\documentclass[linenumbers,preprint,showkeys,preprintnumbers,amsmath,amssymb,superscriptaddress]{revtex4}
\usepackage{amsmath,amssymb,bm,bbm,eucal,graphicx}

\usepackage{graphicx}
\usepackage{graphics}
\usepackage{subfigure}
\usepackage{natbib}
\usepackage{dcolumn}
\usepackage{bm}
\usepackage{txfonts}
\usepackage{color}

\begin{document}

%\setcitestyle{round,authoryear,aysep={},yysep={;}}

\centerline{\Large Spatial heterogeneity in drug concentrations}
\centerline{\Large can facilitate the emergence of resistance to  cancer therapy}

{\small {\vskip 12pt \centerline{Feng Fu$^{a}$, Martin A. Nowak$^{b}$, Sebastian Bonhoeffer$^a$}

\begin{center}
$^a$ Theoretical Biology Group, Institute of Integrative Biology, ETH Zurich, Zurich 8092, Switzerland\\
$^b$ Program for Evolutionary Dynamics, Department of Organismic and
Evolutionary Biology, Department of Mathematics,
Harvard University, Cambridge, Massachusetts 02138, USA\\
\end{center}
}}

\vskip 30pt

\begin{minipage}{142mm}
\begin{flushleft}
{\textbf{Classification:}}\, BIOLOGICAL SCIENCES -- Population Biology

{\textbf{Manuscript information:}}\, 28 pages (including figure legends); 5 figures; 6 supplementary figures; 10 pages supplementary information \\
\vskip 30pt
{\textbf{Corresponding author:}} \\
Feng Fu\\
\small {Institut f. Integrative Biologie, CHN K 18\\
Universit\"{a}tstrasse 16, 8092 Zurich, Switzerland\\
Email: feng.fu@env.ethz.ch\\
Tel: +41 (044) 632 93 03\\
Fax: +41 (044) 632 12 71\\}
\end{flushleft}
\end{minipage}

\clearpage
{
{\bf Abstract:} \, 
Acquired resistance is one of the major barriers to successful cancer therapy. The development of resistance is commonly attributed to genetic heterogeneity. However, heterogeneity of drug penetration of the tumor microenvironment both on the microscopic level within solid tumors as well as on the macroscopic level across metastases may also contribute to acquired drug resistance. Here we use mathematical models to investigate the effect of drug heterogeneity on the probability of escape from treatment and time to resistance. Specifically we address scenarios with sufficiently efficient therapies that suppress growth of all preexisting genetic variants in the compartment with highest drug concentration. To study the joint effect of drug heterogeneity, growth rate, and evolution of resistance we analyze a multitype stochastic branching process describing growth of cancer cells in two compartments with different drug concentration and limited migration between compartments. We show that resistance is more likely to arise first in the low drug compartment and from there populate the high drug compartment. Moreover, we show that only below a threshold rate of cell migration does spatial heterogeneity accelerate resistance evolution, otherwise deterring drug resistance with excessively high migration rates. Our results provide new insights into understanding why cancers tend to quickly become resistant, and that cell migration and the presence of sanctuary sites with little drug exposure are essential to this end.
}


\clearpage

{{\bf Author Summary:} \, 
Failure of cancer therapy is commonly attributed to the outgrowth of pre-existing resistant mutants already present \emph{prior to} treatment, yet there is increasing evidence that the tumor microenvironment influences cell sensitivity to drugs and thus mediates the evolution of resistance \emph{during} treatment. Here, we take into consideration important aspects of the tumor microenvironment, including spatial drug gradients and differential rates of cell proliferation. We show that the dependence of fitness on space together with cell migration facilitates the emergence of acquired resistance. Our analysis indicates that resistant cells that are selected for in compartments with high concentrations are likely to disseminate from sanctuary sites where they first acquire resistance preceding migration. The results suggest that it would be helpful to improve clinical outcomes by combining targeted therapy with anti-metastatic treatment aimed at constraining cell motility as well as by enhancing drug transportation and distribution throughout all metastatic compartments.

\clearpage

%\linenumbers

%Introduction revised
\section{Introduction}

% talk about general aspects of cancer.

Cancer is a common genetic disease that results from accumulated (epi-)genetic changes in tumor cells~\cite{Vogelstein_Science13,Vogelstein_book,Michor_NRC04,Nowak_PNAS04,Bozic_PNAS10}.  Targeted cancer therapy is currently an area of active research~\cite{Sawyers_Nature04}, and is under rapid development~\cite{Druker_NEJM06,Sequist_JCO08,Amado_JCO08,Chapman_NEJM11}. Targeted agents can send cancer into remission, but the response is often short-lived~\cite{Druker_NEJM06,Sequist_JCO08,Amado_JCO08,Chapman_NEJM11,Bozic_Elife13}. The recurrence of cancer, if treated with single agents, is almost certain due to acquired drug resistance~\cite{Gottesman_ARM02,Holohan_NRC13}. Among efforts to understand the rapid acquisition of resistance by cancer cells, particular attention has been paid to the pre-existing resistance arising \emph{prior to} treatment~\cite{Komarova_PNAS05,Iwasa_Genetics06,Michor_CPD06,Durrett_TPB10}.

In parallel, there has been growing interest in studying how the tumor microenvironment influences cell sensitivity to drugs and thus mediates the evolution of resistance during treatment~\cite{Joyce_NRC08,Sierra_DRU05,Quail_NatMed13,Meeds_CCR08,Tredan_NCI07,Meads_NRC09}. Important aspects of the tumor microenvironment include spatial drug gradients and differential rates of cell proliferation~\cite{Tredan_NCI07,Meads_NRC09}. The heterogeneity of this kind can be found both on the microscopic level within solid tumors as well as on the macroscopic level across metastases~\cite{Quail_NatMed13,Tredan_NCI07,Meads_NRC09}. It is not uncommon that incomplete drug distributions within and across metastatic lesions can compromise the efficacy of treatment~\cite{Meads_NRC09,Tredan_NCI07}, which may be in part due to the differences in the ability of the drug  to penetrate different tissues~\cite{Minchinton_NRC06}. As shown in recent experimental and theoretical studies~\cite{Zhang_Science11,Greulich_PRL12,Hermsen_PNAS12,Hallatschek_P12}, drug gradients can help accelerate the evolution of antibiotic resistance. When it comes to anticancer treatment, the consequences of spatial heterogeneity in drug concentrations on the emergence of resistance have yet to be fully addressed.

Amounting evidence suggests metastasis, at least for some cancers, is an early event during primary tumor development~\cite{Pantel_NRCO09,Chambers_NRC02,Joyce_NRC08,Landen_JCO08,Klein_NRC09,Nguyen_NRC09,Kang_CC13}. By activating tissue invasion and metastases~\cite{Hanahan_Cell11}, cancer cells are able to escape from the primary site and disseminate to distant parts of the body, causing life-threatening health problems~\cite{Poste_Nature80,Fidler_NRC03}. At the time of diagnosis and treatment (which  generally occur late in the course of disease), a high proportion of common cancer patients have already had tumor cells disseminated to distant sites for years prior to presentation~\cite{Steeg_NM06,Pantel_NRCO09}.
What is more, clinical outcomes are often complicated by the presence of overt or occult micrometastases in patients~\cite{Pantel_JNCI99,Pantel_NRC04,Braun_NEJM05,Steeg_NM06,Pantel_NRCO09}. It is, therefore, of primary interest to understand the emergence of resistance, particularly in the setting of disseminated cancer.

A recent study reported that circulating tumor cells are detected in 13 out of 36 breast cancer survivors 7-22 years after receiving mastectomy~\cite{Meng_CCR04}. This observation suggests that disseminated cancer, rather than only the primary tumor \emph{in situ}, is actually under stress when treatment is started~\cite{Pantel_NRC04,Steeg_NM06,Pantel_NRCO09}. Moreover, a few studies observed that metastatic cells tend to quickly become chemoresistant~\cite{Kerbel_JCB 94,Shtivelman_On97,Glinsky_CL97,Liang_CCDT02,Smalley_MCT06,Geyer_NEJM06,Meeds_CCR08}, suggesting a positive relationship between metastatic phenotype and the rapid acquisition of drug resistance. It is possible that, with the migration and seeding dynamics~\cite{Chambers_NRC02,Nguyen_NRC09,Steeg_NM06,Sahai_NRC07,Chaffer_Science11,Scott_RSI13,Scott_NRC12}, these metastatic cells located at distant metastatic compartments can re-seed each other, particularly in the presence of tumor sanctuary sites with very little drug exposure. Therefore, it is necessary to explicitly account for the role that the metapopulation structure of metastatic disease plays in the rapid emergence of resistance.

Most recently, in an \emph{in vitro} experiment with metastatic breast cancer cells~\cite{Wu_PNAS13}, it was shown that cell motility and drug gradient of chemotherapy together can lead to fast emerging resistant cells in areas of high concentrations that would otherwise completely inhibit cell growth. This important experimental result begs theoretical questions aimed at revealing the relevant pathways of evolving resistance. Do tumor cells first migrate from low concentration areas and then adapt to high concentration areas? Or alternatively, do they first acquire resistance in low concentration areas and then migrate to and populate areas of exceedingly high concentrations? How do cell motility and drug gradient, or more generally how does the spatial heterogeneity in drug concentrations affect the emergence of resistance? The goal of the present work is to answer these questions and provide qualitative insights by a simple conceptual model. 

Here we focus on the macroscopic level of the tumor microenvironment across metastases, explicitly taking into account the roles of the multi-compartment structure of metastases and cell migration in the emergence of acquired resistance during treatment. Spatial compartments mean different locations that harbor metastatic deposits (i.e., target organs, such as bone marrow, liver, brain and lung)~\cite{Nguyen_NRC09}; migration means dissemination and seeding of cells between metastatic compartments~\cite{Kang_CC13}. We focus on the common population dynamics that govern the evolution of resistance for various cancers differing in their capacity to metastasize. Metastases of solid tumors (such as breast cancer~\cite{Braun_NEJM05} and melanoma~\cite{Chapman_NEJM11}) tend to have well-defined spatial compartments because of low dissemination rate, whereas the compartment structure of liquid cancer (e.g. blood tumors~\cite{Druker_NEJM06}) is diminished by exceedingly high fluidity. By adjusting the migration rate, our model can be suited to study the specific kind of cancer in question.

Clinical observations from multiple sources affirm that an exponential growth model, although remaining an issue of debate~\cite{Laird_BJC65}, is able to adequately describe tumor growth for most cancer patients~\cite{Fournier,Friberg_JSO97,Collins_TNM76,Schwartz_Cancer61,Spratt_Cancer63,Spratt_AS64,Steel_BJC66}. In line with this, we use a stochastic, multi-type branching model to account for the fate of individual cells, particularly these drug-resistant mutations in establishing surviving lineages. Mathematical models of this kind have provided particularly useful insights into understanding evolutionary dynamics of cancer in response to treatment~\cite{Bozic_Elife13,Komarova_PNAS05,Iwasa_Genetics06,Michor_CPD06,Komarova_PLOSONE09,Iwasa_PRSB03,Durrett_TPB10,Danesh_JTB12,Foo_JTO12} (see a review in Ref.~\cite{Foo_JTB14}). A large set of previous models are focused on pre-existing resistance in the primary tumor, arising from neutral evolution prior to treatment~\cite{Iwasa_Genetics06,Durrett_TPB10}. Built on these prior studies, the current work incorporates the compartment structure of metastatic disease and quantifies the role spatial heterogeneity in drug concentrations plays in the evolution of resistance by metastatic cells during treatment.

In our model, cancer cells can migrate from one spatial compartment to another. Spatial heterogeneity in drug concentrations means that there exist sanctuary sites that are not or only partially penetrated by drugs. Therefore, reproductive fitnesses of cells depend not only on their cell types but also on their spatial locations, leading to a rugged fitness landscape (since drug concentrations in different spatial compartments are not necessarily  continuous but discrete in space). As shown in previous studies~\cite{Bozic_Elife13,Komarova_PNAS05,Iwasa_Genetics06}, if the fitness of resistant cells as compared to sensitive cells is neutral or even slightly advantageous in the absence of drugs, the acquisition of resistance not only becomes highly likely but also is accelerated, since there is no selection pressure against resistance. However, it is less clear about the most likely pathway to select for resistance, if resistance mutations incur a fitness cost in the absence of drugs while conferring an advantage over sensitive cells in the presence of drugs. Recent mathematical modeling with laboratory test using mice suggests that resistance carries a fitness cost~\cite{Gatenby_CR09}. In view of this,  the present study is focused on the latter scenario with fitness cost of resistance, although our approach works for any fitness landscape. We show that resistant mutants are unlikely to emerge \emph{in situ} in compartments of high drug concentrations, but arise through the mutation-migration pathway; namely, metastatic cells acquire costly resistance in the sanctuary sites preceding migrating to and thriving in harsh compartments containing high levels of drugs.

\section{Results}

Without loss of generality, let us first study the simplest possible `drug-sanctuary' scenario for treatment failure due to imperfect drug penetration, as illustrated in Fig.~\ref{scheme}. We consider two spatial compartments with dichotomic distributions of drugs: compartment `$0$' can hardly be penetrated by the cancer drug, thereby representing a perfect drug sanctuary site; compartment `$1$' is distributed with adequate amount of drugs that are able to completely wipe out all wild type cells. We assume that one point mutation is sufficient to confer high levels of resistance to the maximum possible concentration of drugs administered during therapy. We denote the genotypes of cells by the number of acquired point mutations: the wild type `$0$' and the resistant type `$1$' (see Materials \& Methods for a detailed description of the model). This minimal model, albeit overly simplified, offers intuitive insights into understanding how spatial drug heterogeneity facilitates the acquisition of \emph{de novo} resistance during cancer therapies. Later on, we will extend this simple model to more realistic cases with multiple cell types and with multiple compartments, where multiple point mutations can successively accrue to confer resistance to exceedingly high drug concentrations~\cite{Wu_PNAS13}.

In the presence of a sanctuary compartment, there exist two competing pathways to lead to the outgrowth of resistance in the compartment of high drug concentrations, as depicted in Figs.~\ref{scheme}b and \ref{scheme}c. One is the ``migration-mutation'' pathway: sensitive cells first emigrate from the sanctuary compartment and then adapt \emph{in situ} to the non-sanctuary compartment with high drug concentrations. The other is the ``mutation-migration'' pathway: sensitive cells first mutate and acquire resistance in the sanctuary compartment, and then migrate to and populate the compartment with high drug concentrations. To understand how the presence of tumor sanctuary sites and cell migration together affect the resistance evolution, we need to determine which pathway provides the more likely path to resistance.

Because cells of the same types have different fitness in the two compartments, we regard migration as a sort of status change in spatial locations. In this way, the two competing pathways can be seen as three-type branching processes, respectively, with different fitness landscapes. Our results are based on multi-type branching processes (Figs.~\ref{scheme}b and \ref{scheme}c)~\cite{Iwasa_PRSB03,Durrett_TPB10,Athreya_book,Goldie_book,Iwasa_JTB04} (see derivation details in the supplementary information). To make progress, let us assume the following rugged fitness landscape owing to the heterogeneity of drug distributions across compartments. In the sanctuary compartment $0$, both sensitive and resistant cells have supercritical replication potential, but resistant cells have slightly lower replication rates than sensitive type due to the cost of the resistant mutation; that is, $b_{00} > d_{00}$, $b_{10} > d_{10}$, and $b_{10} < b_{00}$. In the drug compartment 1, sensitive cells have a subcritical replication potential while resistant cells still have a supercritical replication potential; that is, $b_{01} < d_{01}$, $b_{11} > d_{11}$, and $b_{11} > b_{01}$. 

Although our method works for any mutation rate and migration rate (see Materials \& Methods), we find a simple condition in the limit of low rates of mutation and migration (Fig.~\ref{cpath}). (Such limiting results should hold true in realistic settings of cancer dynamics, given that the point mutation rate of most cancers is estimated to be $10^{-8}\sim 10^{-9}$~\cite{Vogelstein_Science13,Vogelstein_book} and the dissemination rate of pancreatic cancer cells $\sim10^{-7}$~\cite{Haeno_Cell12}.) That is, the mutation-migration pathway is faster than the counterpart, the migration-mutation pathway, to result in resistance in the drug-containing compartment, if the fitness landscape satisfies the following inequality:

\begin{equation}
\frac{b_{00}}{b_{00} - d_{00} - (b_{10} - d_{10})} > \frac{b_{01}}{b_{00} - d_{00} - (b_{01} - d_{01})}.
\label{flc}
\end{equation}

Since most targeted therapies have cytostatic effects, other than the cytotoxic effects commonly seen in traditional chemotherapy~\cite{Leibel}, in this work we assume the drug inhibits cell proliferation and does not affect death rates of cells. The condition above can be greatly simplified when the two compartments provide exactly the same condition for population outgrowth in the absence of drugs. The rugged fitness landscape due to spatial drug heterogeneity, therefore, can be reflected solely by differences in proliferation rates (i.e., $d_{00} = d_{10} = d_{01} = d_{11}$). Moreover, the fitness cost of resistance, $s$, for resistant cells located in the sanctuary compartment $0$ can be parameterised as $b_{10} = (1 - s) b_{00}$, while the fitness cost of sensitivity, $\delta$, for wild type cells in the drug-containing compartment $1$, $b_{01} = (1 - \delta) b_{00}$. Substituting these parameterizations into \eqref{flc}, we arrive at the much simplified condition in terms of $s$ and $\delta$:

\begin{equation}
\delta > \frac{s}{1 + s}.
\label{pathw}
\end{equation}
We immediately observe that this condition is expected to be fulfilled in most cases since potent therapy should be characterized by $\delta >  s$.

Our mathematical framework allows us to calculate the probability of no resistance with respect to time, conditional on non-extinction and starting with a single sensitive cell following each pathway separately (Figure~\ref{cpath}a). We also calculate and compare the average time to resistance (relapse time conditional on non-extinction) following each pathway, as shown in Fig.~\ref{cpath}b. The result demonstrates that the simple condition as given in Eq.~\eqref{pathw} works well in the limit of low migration rate and remains a good approximation for intermediate migration rates. Moreover, if resistant mutation is neutral or even advantageous in the sanctuary, then the mutation-migration pathway is always much faster than the migration-mutation pathway to result in resistance in the drug-containing compartment (Fig.~\ref{cpath}c). 

Another important aspect of tumor microenvironment is characterized by differential rates of proliferation across compartments in the absence of treatment: cells in the sanctuary compartment may have slower replication and turnover rates than these in other compartments~\cite{Tredan_NCI07}. We thus use the parameter $0<\phi\le 1$ to rescale the proliferation and death rates of cells in the sanctuary compartment  0 relative to that in the drug-containing compartment 1: $d_{00}/\phi= d_{10}/\phi=  d_{01} $ and  $b_{01} = (1 - \delta) b_{00}/\phi$ (which implies that in the absence of treatment, cells grow $1/\phi$ times as fast in compartment $1$ as when located in compartment $0$). Although one should refer to the general inequality~\eqref{flc} as the exact condition for the mutation-migration pathway to be predominant, simple algebra shows that the simplified inequality~\eqref{pathw} is still a necessary condition in this case. In fact, reducing the rate of cell proliferation in the sanctuary equally affects the two pathways, delaying the time to resistance (Fig.~\ref{cpath}d). Taken togethers, these results demonstrate that, under a wide variety of conditions (including the ranges of parameter values relevant to cancer), prevailing resistant cells in compartments with high levels of drugs are likely to originate from sanctuary sites, where they acquired resistance preceding migration.

Of interest is to observe the evolutionary process initiated by a single sensitive cell located in the compartment of high drug concentration. We develop a numerical method to show the spatio-temporal dynamics of emerging drug resistance across spatial compartments (see Materials \& Methods). Note that different from the constrained pathways analyzed in Fig.~\ref{cpath}, in this case, migration is allowed to be bi-directional, and both pathways can be at work at the same time. Doing this enables us to study the emergence of resistance in a more natural and realistic setting. Figure \ref{rsnap} shows the joint \emph{probability} distribution of the numbers of resistant cells in the two compartments with respect to time. The skewed distribution in Fig.~\ref{rsnap}a suggests that the sanctuary compartment provides much more favorable condition to evolve resistance than the drug-containing compartment, and thus escaping from the drug-containing compartment to the sanctuary compartment is crucial to this end. 
Constantly seeding the drug-containing compartment with evolved resistant cells tends to make the distribution more balanced (cf Figs.~\ref{rsnap}b and \ref{rsnap}c).
As a result, resistance soon gets established in the drug-containing compartment, and its growth outpaces that in the sanctuary compartment (Fig.~\ref{rsnap}d
), although the chance of having resistance increases steadily with time in both compartments. Similar results are obtained using different initial conditions, despite that resistance evolution is more likely (and sooner) to occur when the sensitive cell is initially placed in the sanctuary compartment than in the drug-containing compartment (cf. Fig. S1 and Fig.~\ref{cpath}). Our results demonstrate that the sanctuary compartment, even though cells are slowly replicating therein, serves as an escape hatch, and indeed is most likely to be the breeding ground of resistance. Therefore, in the presence of sanctuary sites, the overwhelming outgrowth of resistance in the drug-containing compartment is an inevitable outcome due to the mutation-migration pathway. 

Having illuminated the essence of the problem (Figs.~\ref{scheme}-\ref{rsnap}), we now turn to predict outcomes of hypothetical treatments to eradicate (two) metastases in cancer patients (Fig.~\ref{2les}). We assume the total mass of metastasis is relatively small so that treatment starts without any pre-existing resistance. In particular, we assume that the two metastatic lesions differ in size as well as in the level of drug penetration, because of different microenvironments. Specifically, cells in lesion 0 grow much more slowly than in lesion 1 in the absence of drugs, yet the drugs have better penetration of lesion 1 than of lesion 0. To this end, we continuously vary the difference of drug concentrations in the two compartments, $\Delta D = D_1 - D_0$, while keeping the total sum of concentrations constant. Let us now specifically incorporate into the simple model a Hill function that describes concentration-dependent killing efficacy of drugs (see Materials \& Methods).

Without migration and under homogeneous drug concentrations the metastatic cancer almost certainly can be eradicated successfully. However, a worrying situation arises in the presence of a sanctuary with sufficiently low drug concentrations. With increasing difference of drug concentration between the two metastatic compartments, treatment responses exhibit a sharp transition from successful eradication to failed treatment due to acquired resistance (Fig. 4a). Our mathematical framework allows us to calculate the relapse curve, $1 - p_s(t)$, where $p_s(t)$ is the probability of no resistance by time $t$ following treatment, and the (conditional) average time to patient relapse due to acquired resistance (see Materials \& Methods). Figure 4b shows that  the relapse is accelerated by the spatial heterogeneity; the larger $\Delta D$, the faster the relapse. In this simulated hypothetical patient with small lesions, the worst-case scenario is when the drug cannot penetrate lesion 0 at all, and thus the relapse occurs on average approximately $10^3$ days. The relapse could have occurred within weeks if the treatment starts with much bigger lesions (Fig. S2). 

Especially when cells can migrate, resistance results not only from cells originally in the metastatic compartment $0$ but also from these escaping from compartment $1$ . Noteworthy, cell lineages originating from lesion 1 are faster to evolve resistance than these originating from lesion 0 (Fig. 4b). This result is mainly due to the initial condition used: lesion 1 is much larger than lesion 0 that the influx of escaping cells to the sanctuary exceeds the number of cells \emph{in situ}. Since we are considering a branching process, larger population size is more likely to generate resistant mutation. Indeed, as shown in Fig. S2, if the size of lesion 1 is smaller, cells lineages originating from lesion 1 are actually slower to evolve resistance than from lesion 0.

We emphasize that the monotonic decreasing relationship of relapse time with increasing the drug concentration difference, $\Delta D$, is due to: (1) one point mutation is sufficient to confer strong resistance to the maximum possible concentration in compartment 1 ($D_1 = 100$) in this simulated case; (2) compartment 0, which is distributed with less and less amount of drugs with increasing $\Delta D$, provides an increasingly favourable condition for the evolution of resistance and thus renders shorter relapse time, since the mutation-migration pathway is the most contingent pathway for resistance evolution as shown before. Under different assumptions of fitness effects of mutations as described in Eq.~\eqref{DRcurve}, however, multiple point mutations might be required to confer sufficient level of resistance to increasingly high concentrations in compartment $1$ , as $\Delta D$ increases. This variation does not change the general picture about how the presence of sanctuary sites impairs the effectiveness of cancer therapies (Fig. S4a), but the time to \emph{sufficient} level of resistance may well depend on how many point mutations are needed to this end and exhibits an abrupt increase when $\Delta D$ is increased beyond a critical threshold value (the vertical line in Fig. S4b). For very large $\Delta D$, only two-point mutants are able to survive in compartment $1$, wherein the abundances of one-point mutants and sensitive cells are maintained by the immigration-death equilibrium of the branching process. As time passes by, two-point mutants eventually pop up, most likely in the sanctuary compartment $0$, and subsequently immigrate to and populate the compartment $1$.

Taken together, Fig. 4 demonstrates that metapopulation dynamics arising from migration and seeding, together with the presence of sanctuary sites, play an important role in the rapid emergence of resistance. Furthermore, only for migration rates below a certain critical threshold does the spatial heterogeneity in drug concentrations speed up the emergence of resistance. Excessively high migration rates actually slow down resistance emergence (Fig. S3), as the role of compartment structure is diminished by frequent migration and consequently cells are exposed to the drug compartment more often. It is worth noting that there exists an optimum migration rate that leads to the fastest emergence of resistance and that excessively high migration rates actually deter and delay the evolution of resistance (Fig. S3). The results suggest that it may be helpful to improve clinical outcomes by combining targeted therapy with anti-metastatic treatment aimed at inhibiting cell motility as well as by enhancing drug transportation and distribution throughout all metastatic compartments. 

We also extend the simple model to more general cases with multiple cell types and with multiple compartments (see detailed mathematical description in the SI). In this extended model, multiple point mutations can consecutively accrue to confer resistance to increasingly high drug concentrations. In particular, we study and compare the impact of the two different schemes of cell migration, local versus global migration, on the evolution of resistance (Figs. S5 and S6). The numerical results confirm that our conclusions derived from the simple model above remain qualitatively unchanged. Additionally, we observe that strong resistance evolves much faster for local migration than global migration, particularly for sensitive cells located at the sanctuary sites with low drug concentrations (Fig. S5). In other words, sequential local migration over spatial gradient of drug concentration following newly accrued point mutations (i.e., the mutation-migration pathway) plays an important part in leading to rapid selection for high-level resistance. This may be of particular relevance for the evolution of resistance within a solid tumor~\cite{Tredan_NCI07,Wu_PNAS13}. Furthermore, apart from oversimplified migration schemes addressed in the current study, this full model can be readily extend to integrate with a realistic vascular network that regulates the metastatic routes of circulating tumor cells among target organs~\cite{Scott_NRC12}. 

Last but not least, let us demonstrate how the evolution of resistance can be facilitated by the microenvironment within a tumor on the microscopic level~\cite{Tredan_NCI07}. Specifically, we consider a solid tumor consisting of $10^{11}$ cells which is about $4.6$ cm in diameter~\cite{Friberg_JSO97}. Figure~\ref{tm}a shows the schematic representation of tumor microenvironment of cells surrounding a blood vessel located in the centre: the rate of proliferation of tumor cells decreases with increasing distance from the central blood vessel in the absence of treatment. Similarly, the delivery of cancer drugs is also compromised in the presence of treatment, thereby resulting in the spatial drug gradient as illustrated in Fig.~\ref{tm}b. The level of spatial drug heterogeneity is represented by $1/\tau_D$: the larger of this value, the more poorly the drug penetrates distal tumor issues away from the nearest blood vessel. To make progress in our calculations, we artificially divide the tumor into $M=30$ spatial compartments with consecutive concentric circles with equal interval in between from centre to surface. We confirm that dividing more compartments leads to almost the same results as shown here in Figs.~\ref{tm}c and \ref{tm}d. In this simulated example, two point mutations are required to confer full resistance to the maximum drug concentration in the centre. As shown in Fig.~\ref{tm}c, the tumor can be eradicated under perfect drug penetration where the drug is almost homogeneously distributed throughout the entire tumor population (extremely small $1/\tau_D$). In contrast, inadequate penetration of the tumor gives rise to sanctuary sites, these outer compartments that are most distant away from the blood vessel and thus exposed with the least amount of drugs. Therefore, cancer therapy fails certainly with large values of $1/\tau_D$. In line with Fig.~4, relapse occurs sooner with increasing spatial drug heterogeneity, $1/\tau_D$ (Figs.~\ref{tm}d). Moreover, distal cells, although slowly proliferating, are more likely to generate resistance than these proximal cells that are affected most by the drug. These results quantitatively demonstrate that tumor microenvironment mediates cell sensitivity to drugs and thus plays an important role in drug resistance acquired \emph{during} treatments\cite{Tredan_NCI07,Wu_PNAS13}.

\section{Discussion}

Here we study the roles that cell motility and spatial heterogeneity in drug concentrations play in the emergence of acquired resistance to cancer therapy. Cancer cells can migrate from one spatial compartment to another. As compartments may contain different levels of drugs, the cells experience distinct selection pressure for resistance in different compartments. We calculate the probability of resistance and the average time to resistance for sensitive cells originally located in different compartments. We show that the presence of sanctuary sites with poor drug penetration can speed up the emergence of acquired resistance to cancer therapy. Moreover, we show that resistance is unlikely to arise \emph{in situ} within high concentration compartments, but rather that resistance first emerges in sanctuary sites and then spreads to and populates other compartments with high concentrations that would be able to completely inhibit growth of sensitive cells. 
This result is in line with a prior analysis of competing pathways to resistance, based on an ecological source-sink model with logistic population growth~\cite{Hermsen_PRL10}. In spite of that, our study is specifically aimed at understanding the role of tumor microenvironment in the evolution of resistance during treatment. More important, the current work can be used to interpret recent experimental results demonstrating that cell motility and drug gradients can facilitate the emergence of resistance~\cite{Wu_PNAS13}. 

Previous studies both empirically and theoretically reveal that spatial drug gradient can facilitate the evolution of antibiotic resistance~\cite{Zhang_Science11,Greulich_PRL12,Hermsen_PNAS12,Hallatschek_P12}. Although well suited for studying experimental microbial evolution~\cite{Greulich_PRL12,Hermsen_PNAS12}, these models are not directly applicable to the context of cancer; especially solid tumours with intrinsic heterogeneity in their microenvironments warrant a thorough separate investigation~\cite{Joyce_NRC08,Sierra_DRU05,Quail_NatMed13}. Furthermore, in these prior models~\cite{Greulich_PRL12,Hermsen_PNAS12}, compartments are placed in an order with increasing drug concentration, sequential migration occurs only between the two nearest neighbor compartments, and evolution begins within the sanctuary while other compartments are initially void. In contrast, the present mathematical framework takes into account realistic concentration-dependent response for any initial population of cancer cells distributed over multiple metastatic compartments, and allows us to calculate the risk of acquiring resistance as well as to ascertain the timing of relapse. Moreover, our model can be readily extended to incorporate specific migration schemes of cancer cells, \emph{e.g.}, along with a more realistically connected vasculature~\cite{Scott_NRC12}. In parallel, it is worth mentioning that the approach of using partial differential equations to describe spatio-temporal dynamics of cancer evolution~\cite{Anderson_Cell06,Roose_SR07,Anderson_MMB05} sheds a different yet useful insight on the selection of resistance under cancer therapies~\cite{Delitala_JTB12,Lorz_MMNA13}. Taken together, our theoretical results improve our understanding of how metastatic cells can acquire resistance during treatment, especially in the presence of sanctuary sites.

In this study we focus on the evolution of resistance exclusively in the population of metastatic cells that have the same capacity to migrate. It has been shown that migratory cells, though with lower growth potential than non-migratory cells, can be selected for during therapy~\cite{Thalhauser_BD10}. Extending this prior result, our results show that cell motility and the presence of sanctuary sites with little drug exposure are essential for the rapid acquisition of resistance by metastatic cells. Moreover, only for low migration rates below a certain threshold does spatial heterogeneity in drug concentration speed up resistance evolution. Arguably, this finding may help to explain qualitatively the differences of clinical successes in treating liquid cancer (such as chronic myeloid leukemia~\cite{Druker_NEJM06}) and solid tumors (such as melanoma~\cite{Chapman_NEJM11}). Because of limited cell motility, metastases of solid tumor tend to have well-defined spatial compartments (namely, distal lesions), and thus are more prone to drug penetration problems. For this reason, additional attention should be paid to eliminate the sanctuary sites for cancer therapy~\cite{Meeds_CCR08}.

In the current work, we only consider resistance to a single drug, or, to be more precise to treatment with a drug or drug combination to which resistance can be generated by a single mutational event. It is promising for future work to study multi-drug resistance requiring multiple resistance mutations, given that combination therapy is increasingly used in clinical setting~\cite{Baum_Lancet02,Feldmann_CR07}. In addition to the spatial heterogeneity in drug concentrations addressed here, we think that epistatic interactions of resistant mutations to each drug may also be important and deserve further investigation~\cite{Bonhoeffer_Science04}.

Our results suggest that combining targeted therapy with anti-metastatic treatment might help improve clinical outcomes~\cite{Woodhouse_Cancer97,Chambers_ACR00,Wells_TPS13}, especially when treating disseminated cancer. As demonstrated in the present work, inhibition of cell migration between compartments not only suppresses the escape route of sensitive metastatic cells to sanctuary sites and also prevents the dissemination of evolved metastatic cells from sanctuary sites to these compartments with high concentrations, where resistance is strongly selected for. Furthermore, it is desirable to enhance drug transportation and distribution throughout all metastatic compartments, in order to deter the rise of resistance.

The current model is minimalistic, but allows proof of principle. We leave out many important issues, such as cancer stem cells~\cite{Michor_Nature05,Jordan_NEJM06,Visvader_NRC08,Zhou_JTB14}, cellular quiescence and cancer dormancy~\cite{Komarova_PLOSONE07,Aguirre-Ghiso_NRC07}, and the inefficiency of metastatic processes~\cite{Al-Hajj_PNAS03,Wong_CR01,Luzzi_AJP98}. In particular, we have considered the dynamics of the evolution of drug resistance in a situation in which cells are able to move freely from one compartment (tumor) to another. This first approximation, while enlightening, may overestimate some of the dynamics that would occur in a more realistically connected vasculature, especially when the inefficiencies of metastasis, due to filtration and dissemination in the vascular network, are considered~\cite{Scott_RSI13,Scott_NRC12}. With the increasing understanding of the molecular biology of metastasis as well as clinical advances in treating metastasis, we believe that it will become feasible to obtain accurate estimations of key parameters regarding metastatic burden (location and size) and metastatic rates. Then a calibrated model of this sort as introduced here can be used to simulate patient responses \emph{in silico} and predict outcomes of treatments to eradicate the disseminated cancer~\cite{Bozic_Elife13}.

\section{Materials \& Methods}

\paragraph{Minimal model.} We focus the present study on the role that tumor microenvironment plays in the emergence of acquired resistance to potent cancer therapies, where \emph{de novo} mutations are required to confer strong resistance to high drug concentrations. In our model, we explicitly account for the compartment structure of tumor microenvironment as well as the spatial heterogeneity in drug concentrations across compartments. For proof of concept, we focus on the simplest possible case with only two compartments and two types of cells in the main text. Without loss of generality, we assume that drugs have better access to compartment 1 than to compartment 0. In contrast to conventional chemotherapy agents that have cytotoxic effects, most molecularly targeted cancer therapies have cytostatic effects on cancer cells~\cite{Leibel}.  Moreover, it is commonly found that the efficacy of drugs is concentration-dependent in pharmacological kinetics studies ranging from antimicrobial treatment to cancer therapy~\cite{Regoes_AAC04,Hill_IND94,Goutelle_FCP08,Hill_IJC87}. Therefore, it is plausible for us to specifically consider concentration-dependent inhibition of cell replication in response to cancer therapy.

Denote by $D_i$ the drug concentration in compartment $i$.
Upon division, one of the daughter cells can mutate with probability $u$ to become resistant. Denote by $i$  the genotype of a cancer cell if it has acquired $i$ point mutations ($i = 0, 1$). The fitness of a cell depends on its type and spatial location. Specifically, the replication rate of a cancer cell, $b_{ij}$, is determined by its genotype $i$ and spatial location $j$ ($j = 0, 1$) as follows,

\begin{equation}
b_{ij} = \beta_{j}\frac{1 - i s}{1 + \left[ \frac{D_{j}}{\rho^i \text{IC}_{50}}\right]^m}.
\label{DRcurve}
\end{equation}

Here we use a Hill function for the drug response curve~\cite{Rosenbloom_NM12}. $\beta_{j}$ ($\alpha_j$, respectively) is the division rate (the death rate, respectively) of a sensitive cell located in compartment $j$ in the absence of drugs, $s$ is the cost of resistance mutation in the absence of drugs, $\text{IC}_{50}$ is the drug concentration that is needed to inhibit cell growth by one half of its original rate, $\rho$ is the fold increase in $\text{IC}_{50}$ per mutation, and $m$ determines the steepness of the Hill function. Previous studies have fitted empirical data to similar Hill functions as given above (Eq.~\eqref{DRcurve}), to indicate antibiotic resistance and antiviral resistance to various treatment regimens~\cite{Regoes_AAC04,Rosenbloom_NM12}. Although a full characterization of cancer drug resistance in this way has yet to be done, our general results are not dependent on specific parameter choices of $s$, $\rho$, $m$ and $\text{IC}_{50}$. The death rate of a cell with genotype $i$ and in compartment $j$ is unaffected by the presence of drugs and equal to that of sensitive cells irrespective of their genotypes, $d_{ij} = \alpha_j$. The net growth rate is denoted by $r_{ij} = b_{ij} - d_{ij}$. Cells can migrate between the two compartments with rate $v$. The unit of all rates is per cell per day.

\paragraph{Generating function approach.}  Denote by $F_{ij}(\mathbf{X}; t)$ the probability generating function for the lineages at time $t$ initiated by a single $ij$-type cell, where $\mathbf{X} =
 [x_{00}, x_{01}, x_{10}, x_{11}]$ denotes the vector of dummy variables with elements $x_{ij}$ representing each $ij$-type of cells. The backward equations for this branching process are (see SI for how to derive them)
\begin{eqnarray}
\frac{\partial F_{00}}{\partial t} & = &  d_{00} + b_{00}(1-u) F_{00}^2 + b_{00} u F_{00} F_{10} + v F_{01} - (d_{00} + b_{00} + v) F_{00} \nonumber \\
\frac{\partial F_{01}}{\partial t} & = & d_{01} + b_{01}(1-u) F_{01}^2 + b_{01} u F_{01} F_{11} + v F_{00} - (d_{01} + b_{01} + v) F_{01} \nonumber \\
\frac{\partial F_{10}}{\partial t} & = & d_{10} + b_{10} F_{10}^2 + v F_{11} - (d_{10} + b_{10} + v) F_{10}  \nonumber \\
\frac{\partial F_{11}}{\partial t} & = & d_{11} + b_{11} F_{11}^2 +  v F_{10} - (d_{11} + b_{11} + v) F_{11}.
\end{eqnarray} 
The initial condition is given by $F_{ij}(\mathbf{X}; 0) = x_{ij}$.

\paragraph{Probability of acquired resistance.} To extract marginal joint probabilities from generating functions, we use the Cauchy's integral method to replace the task of taking multiple derivatives. Similar methods have been used in the literature~\cite{Antal_JSM10,Conway_PLoSCB11}.
For example, the probability density of the number of resistant cells, $p_{mn}(t)$ in both compartments at time $t$, as shown in Fig. 3, is given by
\[
p_{mn}(t) =\left. \frac{1}{m!}\frac{1}{n!}\frac{\partial F_{01} (1,1,x,y; t)}{\partial x^m \partial y^n}\right |_{x = 0, y =0}.
\]
Using the Cauchy's integral formula, we obtain
\begin{eqnarray}
p_{mn}(t)  & = & \frac{1}{m!n!}\frac{m!n!}{(2\pi i)^2}\oint_c\oint_c\frac{F_{01}(1,1,x,y;t)}{x^{m+1}y^{n+1}}dx dy \nonumber\\
 & = & \frac{1}{4\pi^2}\int_0^{2\pi}\int_0^{2\pi}F_{01} (1,1,e^{i \theta_1},e^{i \theta_2}; t) e^{- i m \theta_1} e^{- i n \theta_2}d\theta_1 d\theta_2
\end{eqnarray}
Applying the trapezoid rule to approximate the double integral, we arrive at:
\[
p_{mn}(t) \approx \frac{1}{N^2}\sum_{i_1 = 0}^{N-1} \sum_{i_2= 0}^{N-1} F_{01} (1,1,e^{i i_1\frac{2\pi}{N}},e^{i i_2\frac{2\pi}{N}}; t) e^{- i m i_1\frac{2\pi}{N}} e^{- i n i_2 \frac{2\pi}{N}}\]
Note that this formula above is essentially equivalent to the discrete Fourier transform of the generating function. To reduce the aliasing effect arising from spectral methods, we set $N = 1000 (\gg 20)$ for the results presented in Fig. 3.

\paragraph{Conditional time to resistance.} We use the conditional probability, $p_s(t)$, of having no resistant cells in both compartments to determine the average time to resistance starting with sensitive cells in either compartment. Specifically, starting with a single sensitive cell in compartment 1, the conditional probability $p_s(t)$ is given by
\begin{equation}
p_s(t) = \frac{F_{00}(1,1,0,0; t) - F_{00}(1,1,0,0; \infty)}{1 - F_{00}(1,1,0,0; \infty)}.
\end{equation}
Then average time to resistance $\bar{T}_r$ (relapse time) can be calculated as follows,
\begin{equation}
\bar{T}_r = \int_0^{\infty} p_s(\tau) d\tau.
\end{equation}

\paragraph{Full model.}
In the supplementary information, we study more general cases where $n$ point mutations are needed to confer full resistance and cancer cells can move between $M$ compartments with restricted local or unrestricted global migration. Drugs are distributed over $M$ spatial compartments according to given levels of spatial concentration heterogeneity. In particular, we consider two different schemes of migration: local migration versus global migration. Local migration means that compartments are situated on a ``ring'' where a cancer cell can only migrate to the two nearest neighbor compartments with equal probability $v/2$. In contrast, global migration means compartments are fully connected where a cancer cell is allowed to migrate from one compartment to any other one with equal probability $v/(M-1)$. This extended model allows us to study how metastatic cancer cells acquire increasing levels of drug resistance as they migrate along a spatial gradient of drug concentration and thrive in  areas of excessively high drug concentrations, as shown in a recent in vitro experiment~\cite{Wu_PNAS13}. 

\section*{Acknowledgements}
We thank Jacob G. Scott and Philip Gerlee for careful reading of and helpful comments on an earlier version of this manuscript. We are greatly indebted to the editors and two anonymous reviewers for their constructive comments which helped us to improve this work. We are grateful for support from the European Research Council Advanced Grant (PBDR 268540), the John Templeton Foundation, the National Science Foundation/National Institute of Health joint program in mathematical biology (NIH grant no. R01GM078986), and the Bill and Melinda Gates Foundation (Grand Challenges grant 37874).

\clearpage

\textbf{Figure legends:} \small
\begin{description}
  \item[Figure~1] Schematic of the simple model. Here we study the scenarios where drug concentration and the rate of cell proliferation can be spatially dependent on the tumor microenvironment and \emph{de novo} resistance mutations are needed to escape potent treatments (\emph{i.e.}, targeted combination therapies). (\textbf{a}) Even if only a small number of cells reside in the sanctuary and/or are slowly replicating, they persistently seed the drug compartment that would have almost certainly become void otherwise. Therefore, resistance in the drug compartment arises through two competing pathways: (\textbf{b}) the ``migration-mutation'' pathway in which sensitive cells first migrate to harsh drug environment and then evolve resistance \emph{in situ}, or (\textbf{c}) the ``mutation-migration'' pathway in which sensitive cells first acquire the resistant mutation in the sanctuary compartment and then migrate into and subsequently populate the drug-present compartment. We show that, under a wide variety of conditions, resistant cells thriving in compartments with high levels of drugs are likely to originate from sanctuary sites, where they acquired resistance preceding migration.
 \item[Figure~2] Competing pathways to selection for resistance. (\textbf{a}) Shown is the probability that no resistant cells are present at time $t$, conditional on non-extinction, for the two respective pathways. Circles are closed-form approximations, and solid lines are obtained by numerically solving the differential equations of the probability generating functions (see SI for details). In (\textbf{b}), the solid lines show the average time to resistance (conditional on non-extinction), following the migration-mutation pathway, as a function of the drug efficacy of growth inhibition of sensitive cells, $\delta$, and with two values of the migration rate, $v$. The dashed horizontal lines are the average time to resistance following the mutation-migration pathway. The vertical line marks the theoretical critical $\delta = s/(1 + s)$, above which the mutation-migration pathway is faster. (\textbf{c}) plots the average time to resistance, following the mutation-migration pathway, as a function of the fitness cost $s$ of resistance in the absence of drugs. Resistance mutations are neutral or advantageous for $s \le 0$, while being costly for $s > 0$. The horizontal line is the average time to resistance following the migration-mutation pathway. The dotted vertical line marks the critical $s$ value, expressed in terms of  $\delta$, $s = \delta/(1-\delta)$. (\textbf{d}) shows the dependence of the average time to resistance on the relative growth rate $r$ of sensitive cells in the sanctuary versus in the drug compartment. The scaling parameter $r \in (0, 1]$ controls the degree to which cells in the sanctuary grow slower than in the drug compartment due to the differences in microenvironment. Reducing the growth rate of cells in the sanctuary prolongs the time to resistance and equally affects the two pathways. Parameters: (\textbf{a} -- \textbf{d}) $b_{11} = 0.45$, $d_{01} = d_{11} = 0.4$, $u = 10^{-4}$; (\textbf{a} -- \textbf{c}) $b_{00} = 0.5$, $d_{00} = d_{10} = 0.4$; (\textbf{c} - \textbf{d}) $b_{01} = 0.35$; (\textbf{a}) $b_{01} = 0.39$, $b_{10} = 0.48$, $v = 10^{-3}$;  (\textbf{b}) $b_{01} = b_{00}(1-\delta)$,  $s = 0.04$, $v = 10^{-4}, 0.05$; (\textbf{c}) $b_{10} = b_{00}(1-s)$, $v = 10^{-4}$; (\textbf{d}) $b_{00} = 0.5 \phi$, $d_{00} = 0.4 \phi$, $s = 0.01$, $v = 10^{-3}$.
\item[Figure~3] Spatio-temporal snapshots of emerging drug resistance. Panels (\textbf{a}) - (\textbf{d}) plot the joint probability density of the numbers of resistant cells in both compartments at time points $t = 10, 50, 100, 200$, respectively, starting with a single sensitive cell placed in the drug compartment. Because the tumor microenvironment mediates the rate of cell proliferation, we assume that resistant cells grow much faster in the drug compartment than in the sanctuary compartment. Panels (\textbf{a}) and (\textbf{b}) show that evolution of resistance  \emph{in situ} in the drug compartment is unlikely, and that the sanctuary compartment provides an escape hatch for sensitive cells originally in the drug compartment to breed resistance. Panels (\textbf{c}) and (\textbf{d}) show that it becomes increasingly likely that not only resistance gets established in the drug compartment due to the constant seeding of resistant cells from the sanctuary compartment, but also its abundance quickly outnumbers that in the sanctuary.
Parameters:
$b_{00} = 0.1$,
$d_{00} = 0.05$,
$b_{01} = 0.38$,
$d_{01} = 0.4$,
$b_{10} = 0.099$,
$d_{10} = 0.05$,
$b_{11} = 0.5$,
$d_{11} = 0.4$,
$u = 10^{-4}$, $v = 10^{-2}$.  
\item[Figure~4] Outcomes of \emph{in silico} treatment to eradicate metastases. The upper row panels (\textbf{a}) show the overall escape probabilities of the two metastatic lesions (dash-dotted curves), and that of the respective each lesion (solid curves), with respect to the increasing difference of drug concentration between the two metastatic compartments, $\Delta D = D_1 - D_0$, and for varying migration rates, $v$. Corresponding to (\textbf{a}), the lower row panels (\textbf{b}) show the average time to resistance (conditional on non-extinction), as a function of the level of heterogeneity in drug concentrations, $\Delta D$. When treatment starts, a hypothetical cancer patient has two metastatic lesions: lesion $0$ has $N_0 = 10^3$ cells and lesion $1$ has $N_1 = 10^8$ cells. We also assume that cells in metastatic compartment $0$ ($\beta_0 = 0.05$, $\alpha_0 = 0.04$) grow much slower than in compartment $1$ ($\beta_1 = 0.5$, $\alpha_1 = 0.4$), in the absence of drugs. We assume the drugs have poorer penetration of lesion 0 than of lesion 1, and vary the difference of drug concentration from 0 ($D_0 = D_1 = 50$) to 100 ($D_0 = 0, D_1 = 100$). In this simulated case, only one point mutation is needed to confer strong resistance to high drug concentrations. Parameters: $\text{IC}_{50} = 50$, $m = 2$, $\bar{D} = (D_0 + D_1)/2 = 50$, $s = 0.01$, $\rho = 5$, $u = 10^{-9}$, $v = 0, 10^{-4}, 10^{-3}, 10^{-2}$.

\item[Figure~5] Tumor microenvironment facilitates the emergence of high levels of drug resistance. For simplicity, we specify the microenvironment of tumor cells in relation to their distance to the nearest blood vessel, $x$. Shown in (\textbf{a}) is the schematic section view of  an ``onion-structured'' solid tumor with the nearest blood vessel located in the center. In the absence of drugs, both proliferation and turnover rates of cancer cells decrease with $x$~\cite{Tredan_NCI07}: birth rate $\beta(x) = b_0\exp(-x/\tau_g)$ and death rate $\alpha(x) = d_0\exp(-x/\tau_g)$, where the parameter $\tau_g$ is the characteristic length scale of spatial decay in proliferation and turnover rates. Similarly, the spatial density distribution of tumor cells is assumed to exponentially decay with $x$ and proportional to $\exp(-x/\tau_c)$, where $\tau_c$ is the characteristic length scale of decrease in cell density. Panel (\textbf{b}) plots the spatial drug gradient mediated by the tumor microenvironment~\cite{Tredan_NCI07}: $D(x) = D_0\exp(-x/\tau_D)$, where $D_0$ is the maximum possible concentration in the center and $\tau_D$ is the characteristic length scale of spatial decay of drug concentrations with respect to the distance to the blood vessel, $x$. Panels (\textbf{c}) and \textbf (d) show the escape probability and the average time to resistance (conditional relapse time to acquisition of two point mutations) as a function of the level of spatial drug heterogeneity, $1/ \tau_D$. Using a series of consecutive concentric circles with equal interval in between, we artificially divide the tumor into $M$ compartments (similar to contour lines shown in \textbf{b}). Cells in the same compartment are regarded as homogeneous subpopulations. Tumor cells can migrate to the two nearest neighbouring compartments with equal probability $v/2$. Acquisition of two point mutations is needed to survive in the centre with the maximum drug concentration, $D_0$. Parameters: tumor size $N = 10^{11}$, $\tau_g = \tau_c = 1\text{cm}$, $D_0 = 500$, $b_0 = 0.5$, $d_0 = 0.4$, $M = 30$, $n = 3$, $\text{IC}_{50} = 50$, $m = 2$, $s = 0.01$, $\rho = 5$, $u = 10^{-9}$, $v = 2\times10^{-4}$.
%Other supporting components of tumor microenvironment such as stromal cells and extracellular matrix are not shown. 

\end{description}

\newpage
\textbf{Supporting Information Legends:} \small
\begin{description}

  \item[Figure~S1] Spatio-temporal snapshots of emerging drug resistance. Panels (\textbf{a}) - (\textbf{d}) plot the joint probability density distribution of the numbers of resistant cells in both compartments at time points $t = 10, 50, 100, 200$, respectively, starting with a single sensitive cell placed in the sanctuary compartment. Parameters are the same as in Fig. 3.
  
   \item[Figure~S2] The average time to resistance as a function of the size of lesion $1$, rescaled by the mutation rate, $u N_1$. Relapse occurs sooner with bigger tumor size at the start of therapy. Relapse is destined to happen in the presence of drug sanctuary and cell motility. The situation is even worse for really big tumor sizes at the start of therapy; relapse can happen within weeks. Parameters: $D_0 = 0, D_1 = 100$, $v = 10^{-4}$, and other parameters are the same as in Fig. 4.
   
   \item[Figure~S3] Optimal migration rate. Parameters: $D_0 = 0, D_1 = 100$, and other parameters are the same as in Fig. 4.
    
   \item[Figure~S4] The relapse time depends on the number of point mutations that is needed to confer sufficient levels of resistance to increasingly high drug concentrations. (\textbf{a}) shows the escape probability as a function of the difference in drug concentrations between the two metastatic compartments, $\Delta D$. (\textbf{b}) shows how the average time to resistance changes with increasing $\Delta D$. The vertical line marks the critical value of $\Delta D$ above which two point mutations are required to confer sufficient levels of resistance to increasingly high concentrations in compartment $1$ while compartment $0$ is the sanctuary containing lower level of drugs. Parameters: $\rho = 3.5$, $v = 10^{-4}$, and other parameters are the same as in Fig. 4.

  \item[Figure~S5] The role of cell motility in the emergence of full resistance under local and global migrations. Shown are the escape probabilities and the average conditional time to resistance for a single sensitive cell initially placed in each compartment, with increasing migration rate, $v$. The spatial heterogeneity in drug concentrations is realized by using a rescaled Normal distribution with a peak in the central compartment, and the level of heterogeneity is denoted by the standard deviation, $\sigma$, of the concentration distribution over compartments.
Parameters: $M = 20$, $n = 5$, $\text{IC}_{50} = 100$, $m = 2$, $\rho = 1.1$, $\bar{D} = 50$, $s = 0.01$, $b_0 = 0.2$, $d_0 = 0.1$, $\mu = 10^{-4}$, $\sigma = 24.7$.

 \item[Figure~S6] The impact of the spatial heterogeneity in drug concentrations on the emergence of full resistance under local and global migrations. Shown are the escape probabilities and the average conditional time to resistance for a single sensitive cell initially placed in each compartment, with increasing spatial heterogeneity, $\sigma$. The spatial heterogeneity in drug concentrations is realized by using a rescaled Normal distribution with a peak in the central compartment, and the level of heterogeneity is denoted by the standard deviation, $\sigma$, of the concentration distribution over compartments.
Parameters: $M = 20$, $n = 5$, $\text{IC}_{50} = 100$, $m = 2$, $\rho = 1.1$, $\bar{D} = 50$, $s = 0.01$, $b_0 = 0.2$, $d_0 = 0.1$, $\mu = 10^{-4}$, $v = 0.01$.
\end{description}

\newpage

\begin{figure}[!ht]
\begin{center}
\includegraphics[width=18cm]{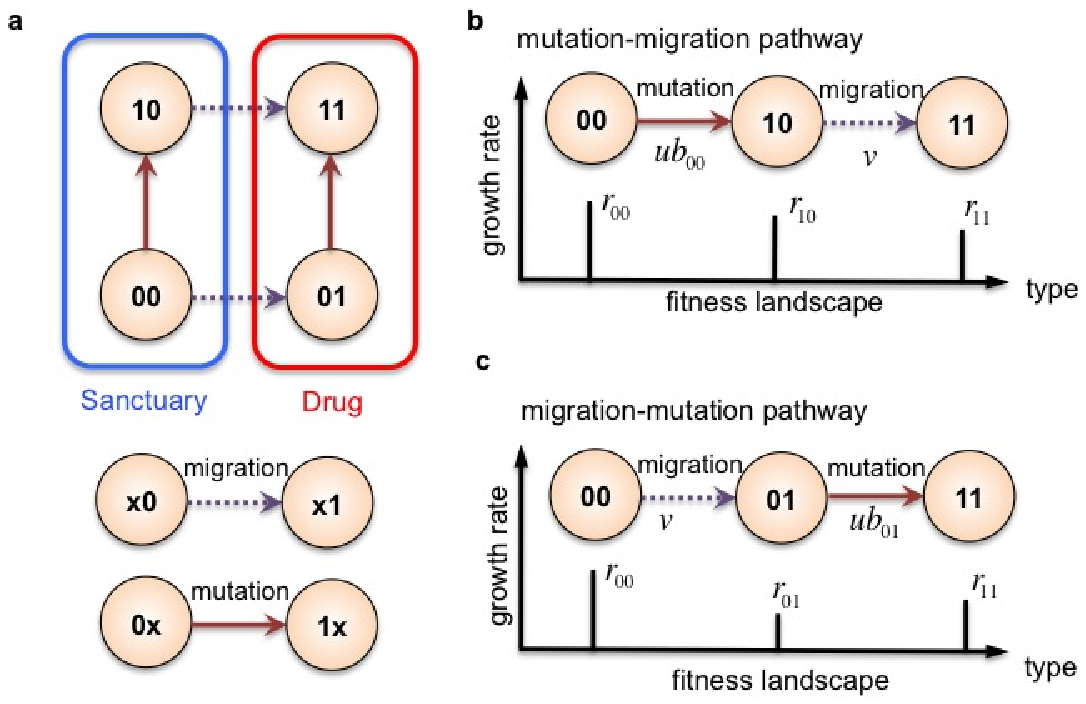}
\end{center}
\caption{}
\label{scheme}
\end{figure}

\begin{figure}[!ht]
\begin{center}
\includegraphics[width=\columnwidth]{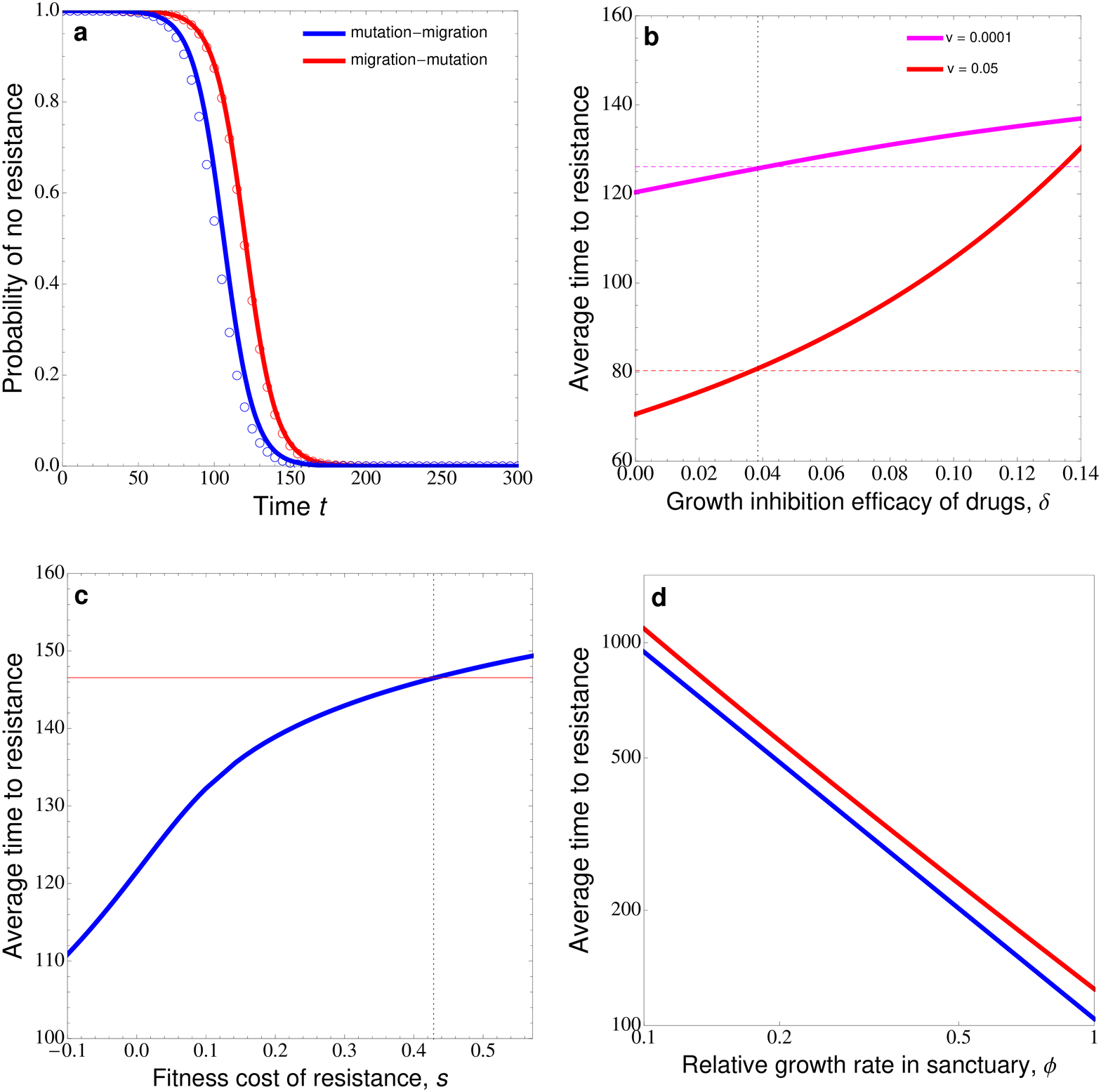}
\end{center}
\caption{}
\label{cpath}
\end{figure}

\begin{figure}[!ht]
\begin{center}
\includegraphics[width = \columnwidth]{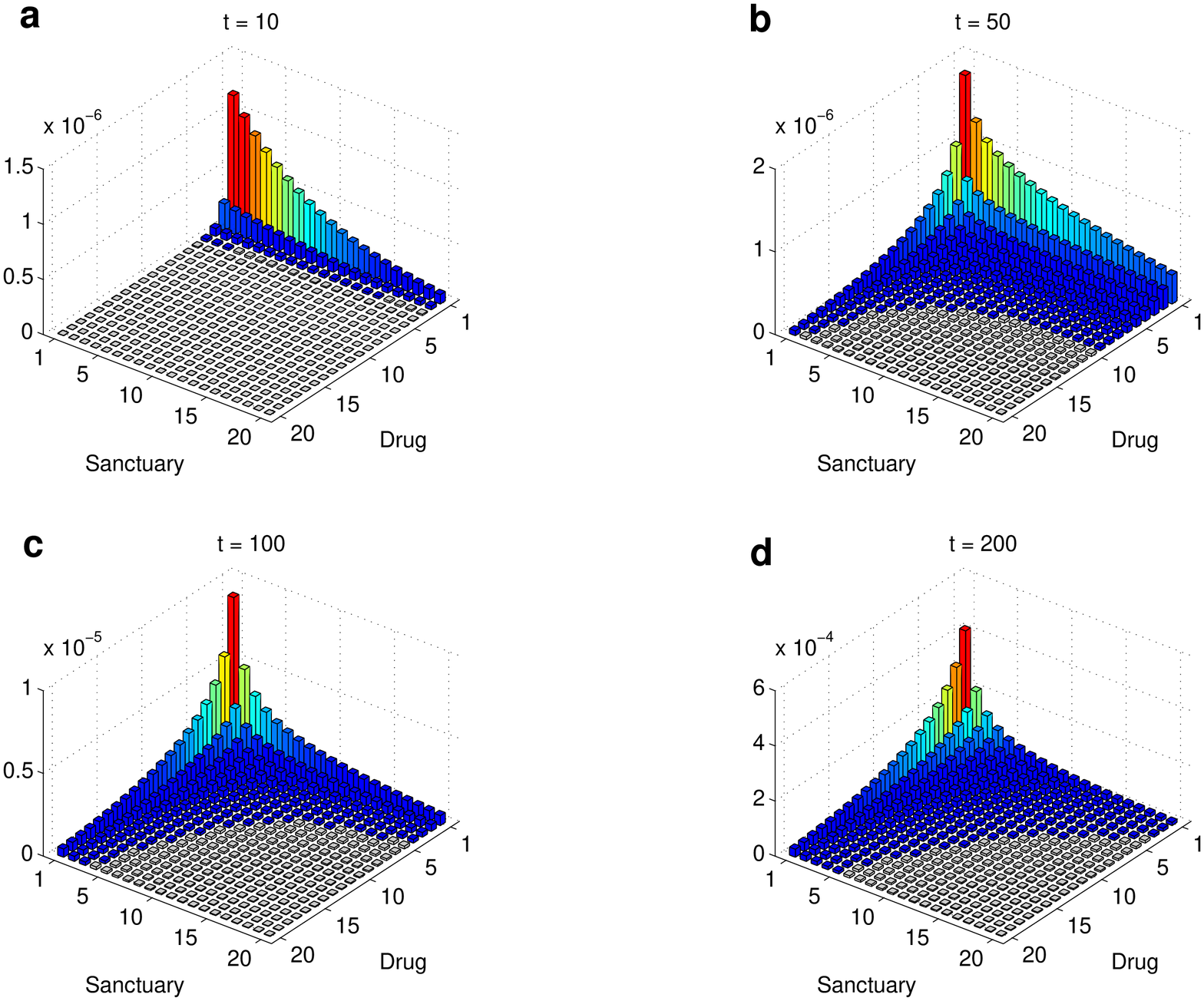}
\end{center}
\caption{}
\label{rsnap}
\end{figure}
\clearpage

\newpage
\begin{figure}[!ht]
\begin{center}
\includegraphics[width=20cm]{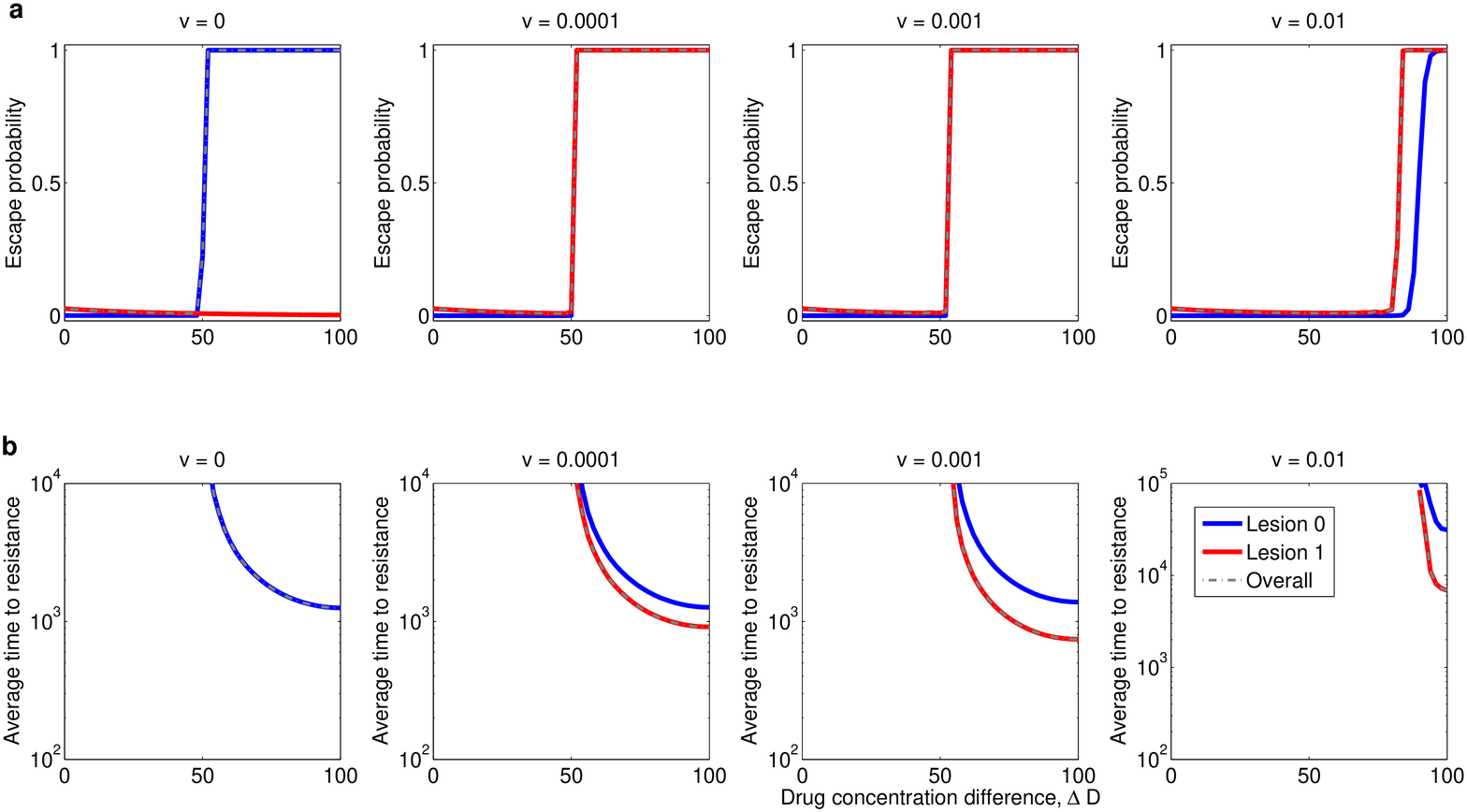}
\end{center}
\caption{
}
\label{2les}
\end{figure}

\newpage
\begin{figure}[!ht]
\begin{center}
\includegraphics[width=20cm]{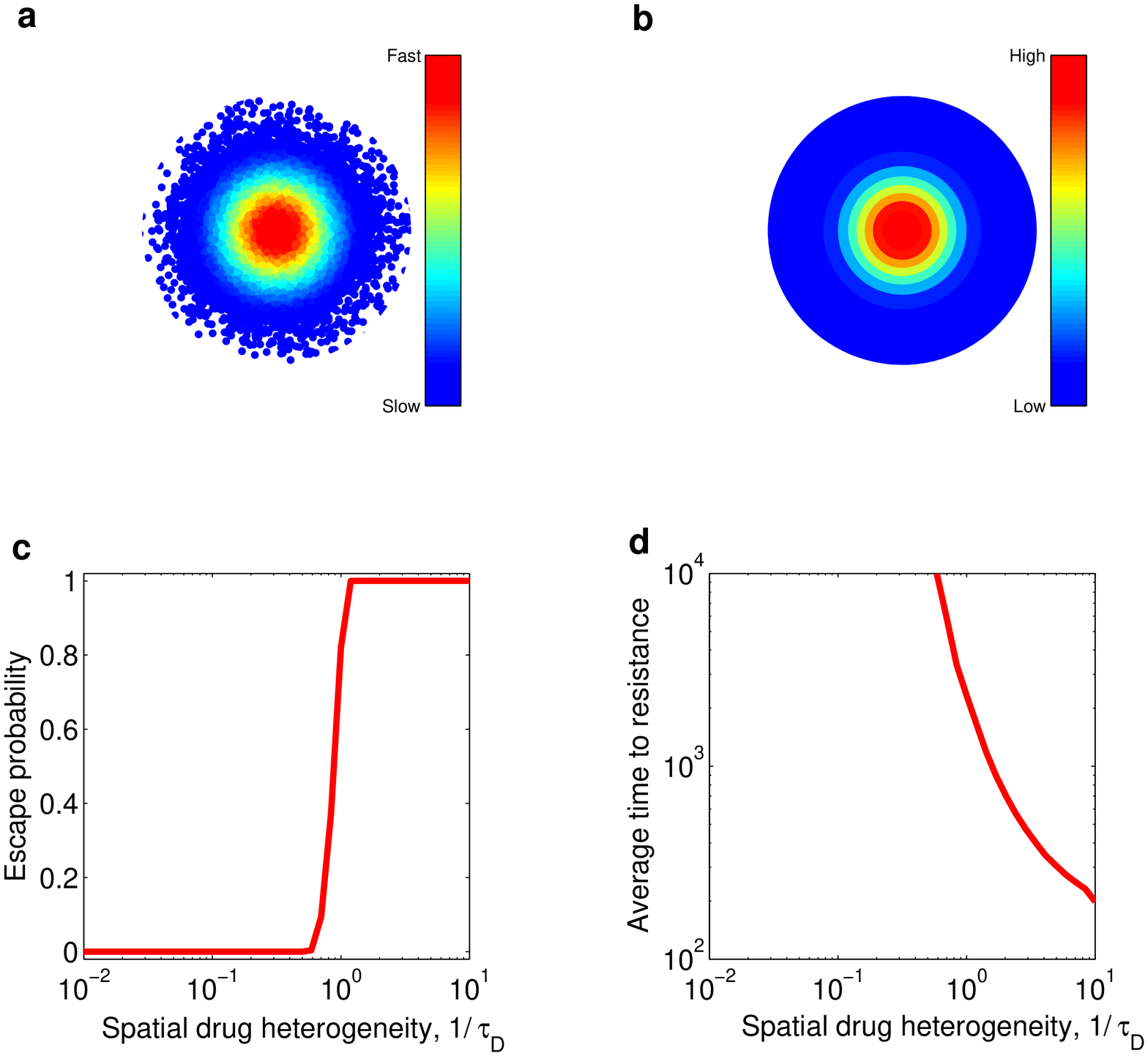}
\end{center}
\caption{
}
\label{tm}
\end{figure}

\newpage
\renewcommand{\figurename}{FigS.}
\setcounter{figure}{0}

\begin{figure}[!ht]
\begin{center}
\includegraphics[width=18cm]{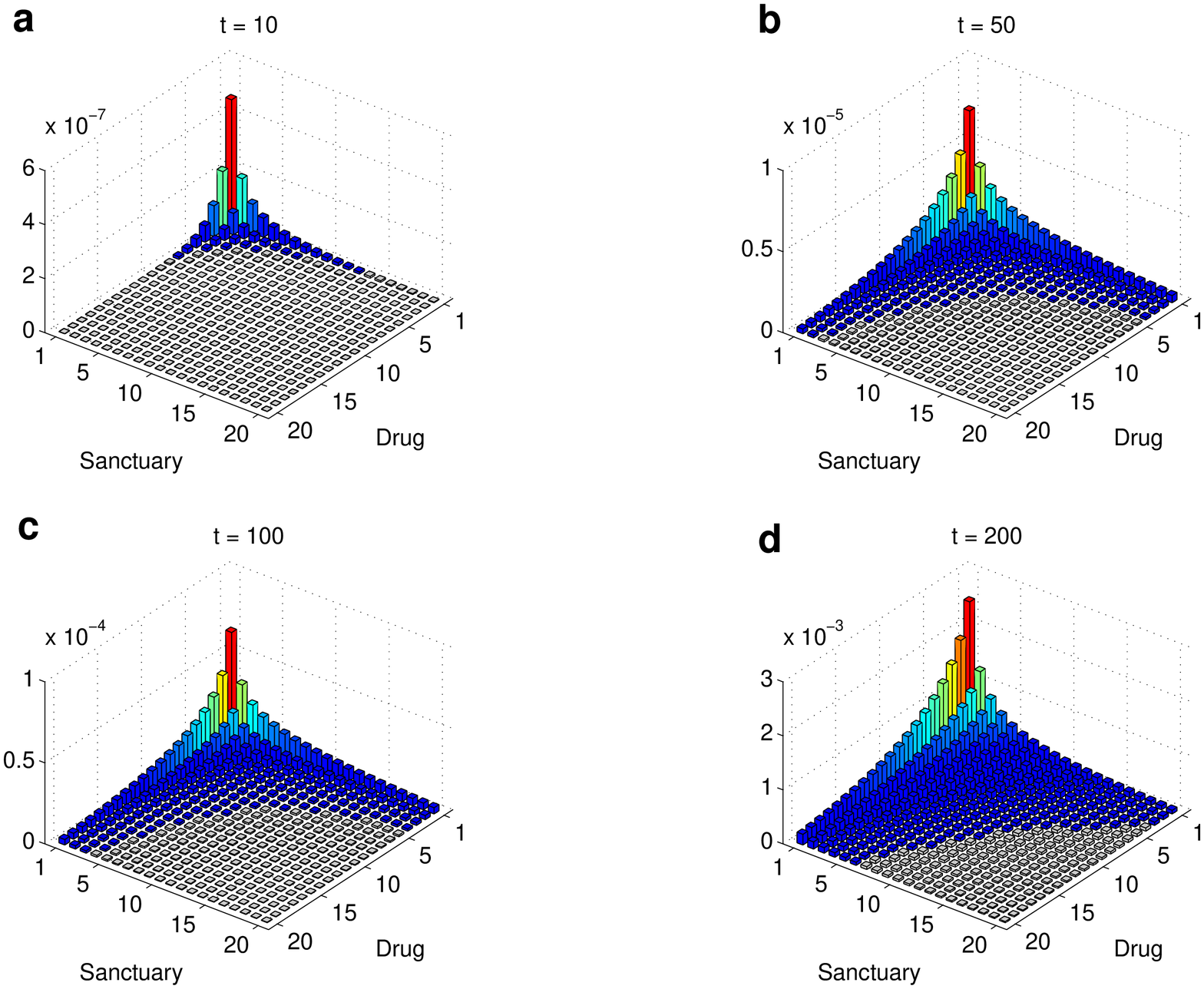}
\end{center}
\caption{}

\end{figure}

\begin{figure}[!ht]
\begin{center}
\includegraphics[width=18cm]{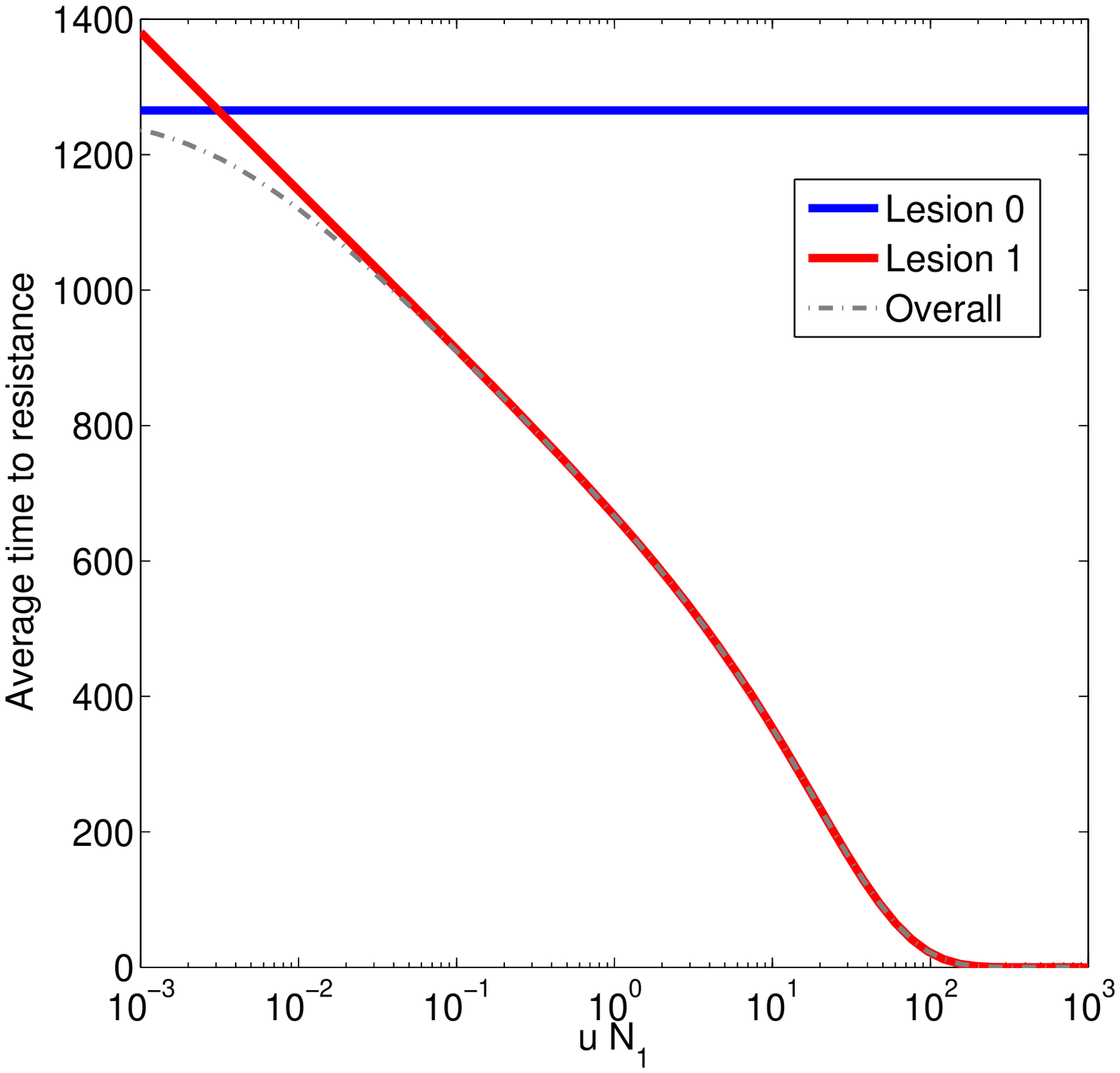}
\end{center}
\caption{}

\end{figure}

\begin{figure}[!ht]
\begin{center}
\includegraphics[width=18cm]{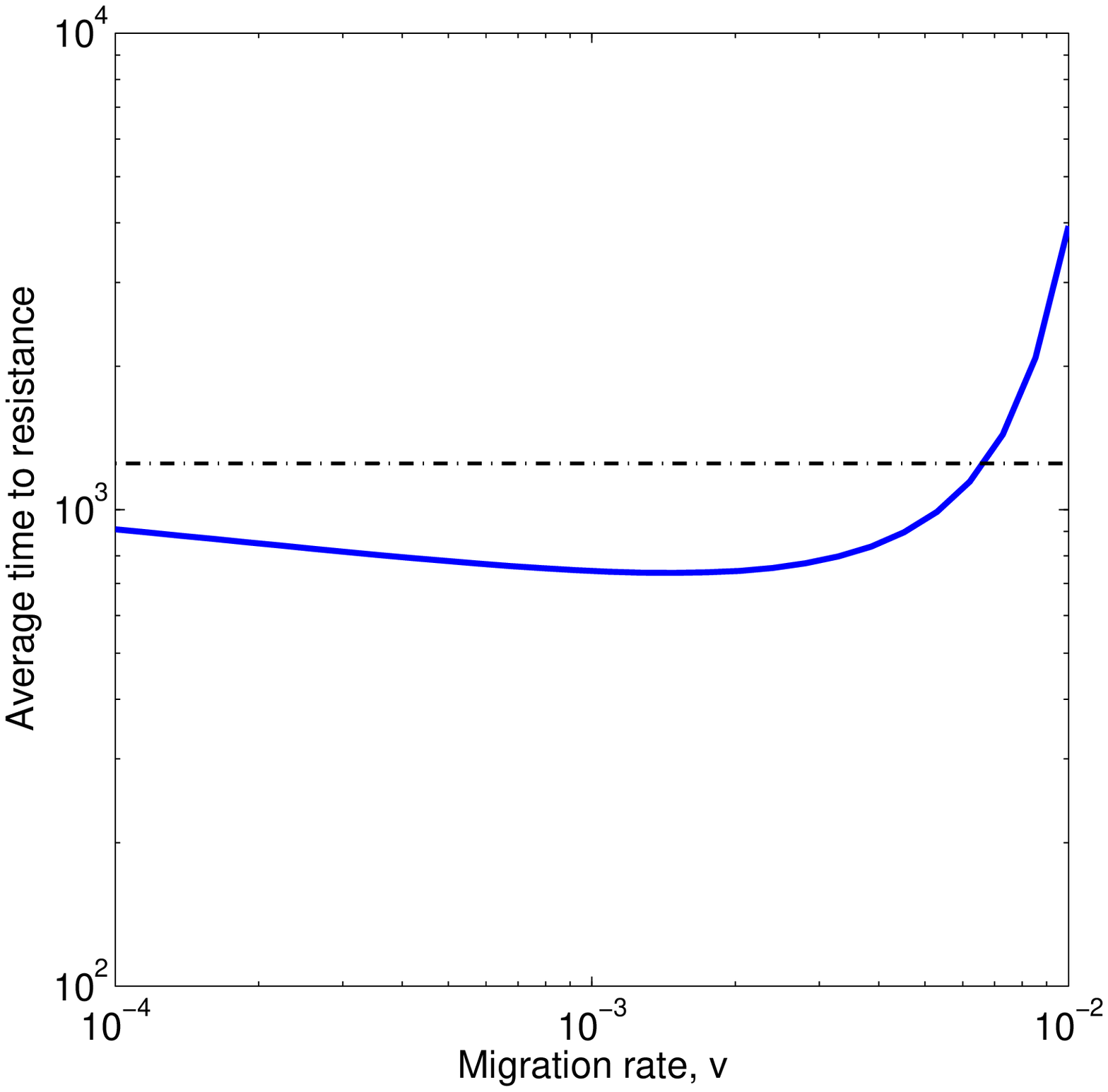}
\end{center}
\caption{}

\end{figure}

\begin{figure}[!ht]
\begin{center}
\includegraphics[width=16cm]{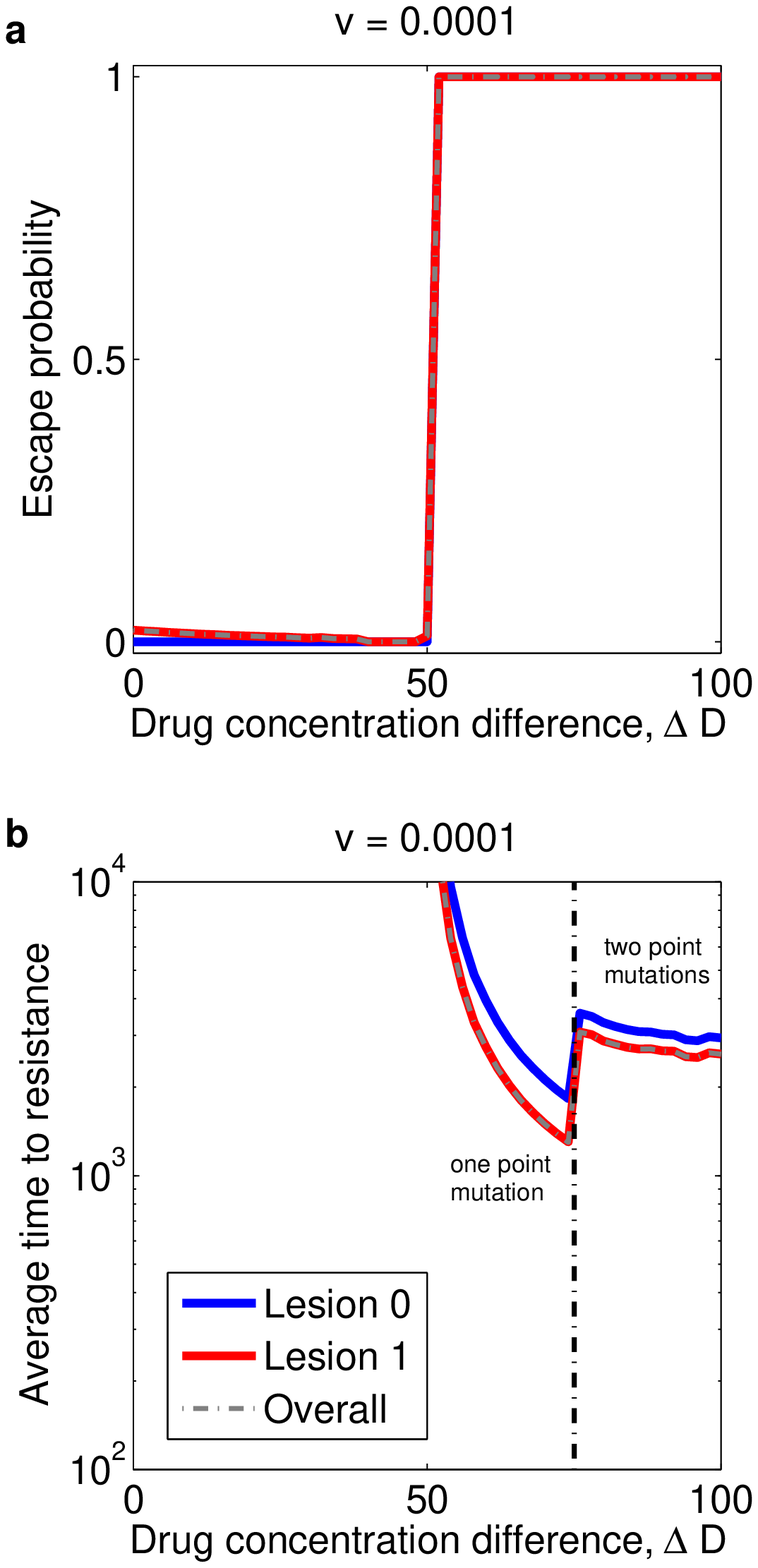}
\end{center}
\caption{}

\end{figure}

\begin{figure}[!ht]
\begin{center}
\includegraphics[width=18cm]{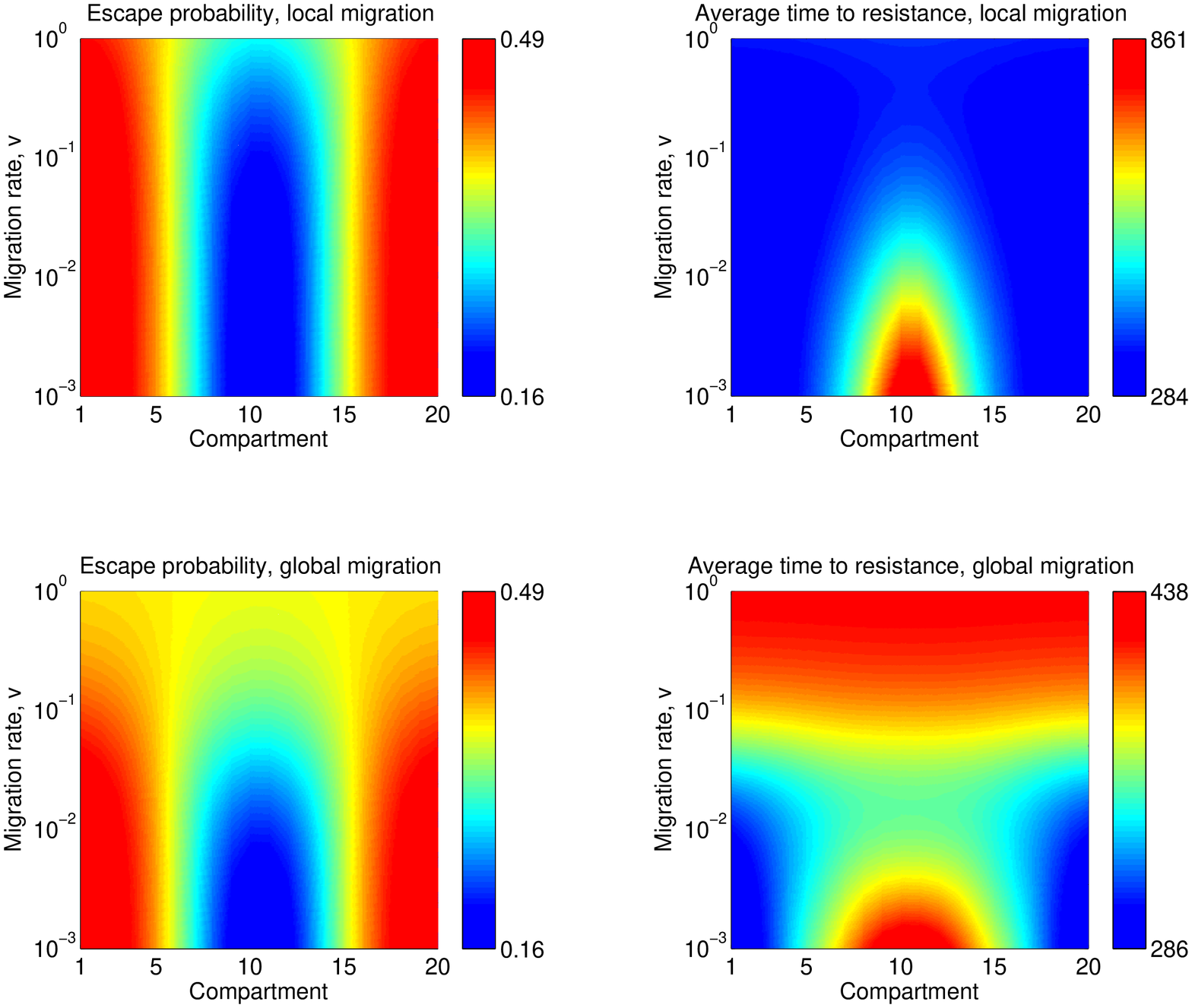}
\end{center}
\caption{}
\label{lgm1}
\end{figure}

\begin{figure}[!ht]
\begin{center}
\includegraphics[width=18cm]{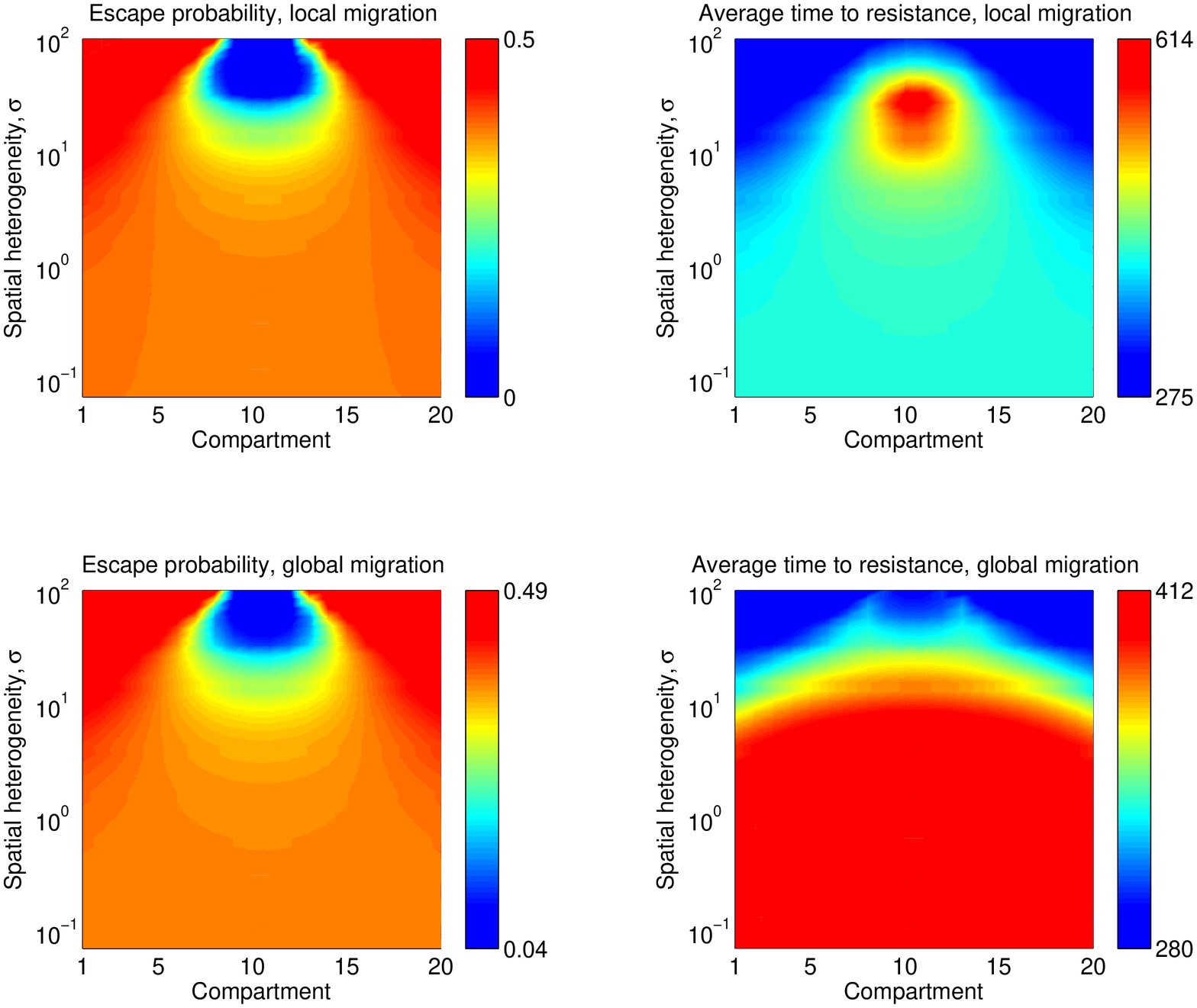}
\end{center}
\caption{}
\label{lgm2}
\end{figure}

\clearpage

\appendix

\title{\Huge{Electronic Supplementary Material for\\
``Spatial heterogeneity in drug concentrations can facilitate the emergence of resistance to  cancer therapy''}}
\author{Feng Fu,\,\,Martin A. Nowak,\,\,Sebastian Bonhoeffer}

\date{\today}
%\tableofcontents

\maketitle

\section{Full model}
For completeness, here we present in details the model description and the analytical methods. In this paper, we use a simple spatial model to study the evolution of resistance to cancer therapy, particularly in the presence of sanctuary sites with low or no drug concentration. Our approach is based on branching process theory. The present study focuses only on (single-) drug resistance arising during treatment, rather than pre-existing resistance prior to treatment. When cancer patients with metastatic diseases are administered potent drugs, it is often the case that the drug is not uniformly distributed throughout the body compartments of the cancer patient due to various factors such as drug penetration problems. This leads to non-homogeneous drug concentrations across different body compartments, which may harbor small metastatic populations. These compartments with low drug concentrations, which are insufficient to fully inhibit the replication of cancer cells, become the sanctuaries from which drug resistant mutants can emerge. Through migration these resistant mutants can subsequently populate compartments with high drug concentration, where the sensitive type cannot survive. By such intuitive reasoning, we hypothesize that the spatial heterogeneity in drug concentrations can speed up the evolution of acquired resistance to cancer therapy. In the rest of this supplementary information, we shall derive mathematical results to investigate this hypothesis. Our theoretical results shed new light onto the emergence of drug resistance and treatment failure of cancer patients with metastatic diseases. 

Let us consider a spatial compartment model of metastatic diseases. There are $M$ compartments in total and the drug concentration in each compartment $i$ is denoted by $D_i$. We restrict our analysis on spatially heterogeneous drug concentrations, but without time-changing concentration fluctuations. Metastatic cancer cells have already acquired motility and are assumed to migrate between compartments with rate $v$. The effect of the drug can either be cytostatic inhibiting cell growth and division or be cytotoxic killing cells directly, or both. Without loss of generality, we assume here that the drug inhibits cell growth. Reproduction is subject to mutation. Upon division, one of the two daughter cells mutates with rate $u$. By acquiring step-wise point mutations, the mutant becomes more and more adapted to drug environments and can eventually survive in high-concentration compartments. Denote by $i$  the genotype of a cancer cell if it has acquired $i$ point mutations. The replication rate of a cancer cell, $b_{ij}$, is determined by its genotype $i$ and spatial location $j$ as follows,
\begin{equation}
b_{ij} = \beta_j\frac{1 - i s}{1 + \left[ \frac{D_{j}}{\rho^i \text{IC}_{50}}\right]^m}.
\label{DRcurve}
\end{equation}
Here we use a Hill function for the drug response curve. $\beta_j$ is the division rate of a wild type cell in compartment $j$ in the absence of drug, $s$ is the cost of mutation in the absence of drug, $\text{IC}_{50}$ is the drug concentration that is needed to inhibit cell growth by one half of its original rate, $\rho$ is the fold increase in $\text{IC}_{50}$ per mutation, and $m$ determines the steepness of the Hill function. We note that drug mutations are deleterious in the absence of drugs. Thus the replication rate of a mutant with genotype $i$ and in compartment $j$ is $(1 -  i s)\beta_j$ in the absence of drugs, whereas its $\text{IC}_{50}$ is increased to $\rho^i \text{IC}_{50}$ in the presence of drugs. Throughout this supplementary information, unless stated otherwise, the death rate of a cancer cell with genotype $i$ within compartment $j$, $d_{ij}$, is all the same across compartment and equal to that of the wild type in the absence of drugs, $d_{ij} = \alpha_j = d_0$. 

The migration pathway is specified by the underlying connections between compartments. For simplicity, we consider two different schemes of migration: local migration versus global migration. Local migration means that compartments are situated on a ``ring'' where a cancer cell can only migrate to the two nearest neighbor compartments with equal probability $v/2$. In contrast, global migration means compartments are fully connected where a cancer cell is allowed to migrate from one compartment to any other one with equal probability $v/(M-1)$. We restrict our subsequent analysis to these two foregoing migration patterns, though it is straightforward to extend it to more complicated migration patterns in our mathematical model.

In what follows, we shall use a continuous-time multi-type branching process to describe resistance evolution. 

\section{Analysis}

As for multi-type branching process,  we first need to construct the appropriate type space using the combination of genotype $g_i \in \{0, 1, \cdots, n-1\}$ and compartment location $l_j \in \{0, 1, \cdots, M-1\}$. Thus a cell's type can be denoted by two-bit strings $g_il_j \in \{0, 1, \cdots, n-1\}\times \{0,1, \cdots, M-1\}$ and thus there are $n\times M$ different types. For example, a cell with ``$11$''-type means that this cell has accumulated $1$ point mutation and is located in compartment $1$. Therefore, a cell's type, notated by its genotype and spatial location, can change as results of genetic mutation or migration.

Next let us introduce the probability generating functions. Denote by $F_{ij}(\mathbf{X}; t)$ the probability generating function for the lineages at time $t$ initiated by a single $ij$-type cell, where $\mathbf{X} =
 [x_{00}, \dots, x_{n-1,M-1}]$ denotes the vector of dummy variables with elements $x_{ij}$ representing each $ij$-type of cells. During an infinitesimal time interval $\Delta t$, with probability $b_{ij} (1-u) \Delta t$ an $ij$-type cell gives birth to two identical daughter cells, with probability $b_{ij} u  \Delta t$ it gives rise to an identical offspring and a mutated offspring, with probability $v\Delta t$ it migrates to another compartment, with probability $d_{ij} \Delta t$ it dies, and with probability $1 - (b_{ij} + d_{ij} + v)\Delta t$ nothing changes. We can use the probability generating function approach to account for what can happen to an $ij$-type cell  during an infinitesimal time interval $\Delta t$,
 \begin{equation}
 f_{ij}(\mathbf{X}; \Delta t) = d_{ij}  + b_{ij} (1-u) \Delta t x_{ij}^2 + b_{ij} u  \Delta t x_{ij} x_{(i+1)j} + \frac{v}{M-1} \Delta t \sum_{k\neq j} x_{ik} + [1 - (b_{ij} + d_{ij} + v)\Delta t] x_{ij}.
 \end{equation}
Using the multiplicative property of branching process (the independency of each individual lineage), the generating function $F_{ij}(\mathbf{X}; t + \Delta t) $ satisfies the recursive backward equation,
\begin{equation}
F_{ij}(\mathbf{X}; t + \Delta t) = f_{ij}(\mathbf{F}; \Delta t),
\end{equation} 
where $\mathbf{F}$ is the vector of the probability generating functions with elements $F_{ij}(\mathbf{X}; t )$. We obtain the following backward equation for global migration ($0\le i < n-1, 0\le j \le M-1$). 
\begin{eqnarray}
\frac{\partial F_{ij}(\mathbf{X}; t )}{\partial t} & = & \lim_{\Delta  t \to 0} \frac{f_{ij}(\mathbf{F}(t); \Delta t) - F_{ij}(\mathbf{X}; t )}{\Delta t} \nonumber \\
& = & d_{ij}  + b_{ij} (1-u) F_{ij}^2 + b_{ij} u F_{ij}F_{(i+1)j} + \frac{v}{M-1}  \sum_{k\neq j} F_{ik}  - (b_{ij} + d_{ij} + v) F_{ij}.
\label{GME}
\end{eqnarray}
Since we do not consider more than $i = n-1$ point mutations, for $i = n - 1$ we have 
\begin{equation}
\frac{\partial F_{ij}(\mathbf{X}; t )}{\partial t} = d_{ij}  + b_{ij} F_{ij}^2 + \frac{v}{M-1}  \sum_{k\neq j} F_{ik}  - (b_{ij} + d_{ij} + v) F_{ij}.
\end{equation}
The initial condition for these ODE's is $F_{ij}(\mathbf{X}; 0) = x_{ij}$ for $i = 0, \dots, n-1$ and $j = 0, \dots, M-1$. For any given profile of the initial numbers of cells, $\mathbf{N} = \{N_{ij}\}$, the corresponding probability generating function $F_{\mathbf{N}}$ is a power compound of $F_{ij}$
\begin{equation}
F_{\mathbf{N}} = \prod_{i, j} F_{ij}(\mathbf{X}; t)^{N_{ij}}.
\end{equation}

Analogously with Eq.~\eqref{GME}, we can derive the backward equation for local migration,
\begin{equation}
\frac{\partial F_{ij}(\mathbf{X}; t )}{\partial t} = d_{ij}  + b_{ij} (1-u) F_{ij}^2 + b_{ij} u F_{ij}F_{(i+1)j} + \frac{v}{2} F_{i [(j-1+M)\%M]} +  \frac{v}{2} F_{i [(j + 1)\%M]} - (b_{ij} + d_{ij} + v) F_{ij}.
\label{LME}
\end{equation}
In the above equation, the subscripts $[(j-1+M)\%M]$ and $[(j + 1)\%M]$ account for the two neighbor compartments the cell may migrate to.

In the following unless stated otherwise, we have $d_{ij} = d_0$ for all types of cells, while $b_{ij}$ depends on the drug concentration in compartment $j$ as in Eq.~\eqref{DRcurve}. The fate of a $ij$-type cell can either go extinct in a relatively short time or successfully establish surviving lineages with expanding populations. The extinction probability, which is the opposite of escape probability, can be obtained by solving the fixed points of the ODE's given above. We can also calculate the probability that there is no evolved fully resistant strain in any compartment by time $t$, conditional on non-extinction.

For arbitrary number of genotypes and arbitrary number of compartments, we cannot find closed form probability generating functions for these nonlinear ODE's. To circumvent this difficulty, we rely on numerical ODE solvers to obtain the quantities of interest. For instance, if we solve the above nonlinear ODE's with the initial condition $x_{ij} =0$, we can obtain the extinction probability $F_{ij}(\mathbf{0}; t)$ for each $ij$-type cell at time $t$. For $t\to \infty$, $F_{ij}(\mathbf{0}; t)$ converges to the steady equilibrium value (i.e., the smallest of all admissible fixed points). Using the initial condition $\mathbf{X}_{noR} = [x_{ij}]$ where $x_{ij} =1$ for $i< n-1$ and $x_{ij}=0$ for $i = n-1$, the conditional probability $P_{noR}(t)$ that there is no evolved fully resistant strain in any compartment by time $t$, starting with an $ij$-type cell, can be calculated as,
\begin{equation}
P_{noR}(t) = \frac{F_{ij}(\mathbf{X}_{noR}; t) - F_{ij}(\mathbf{0};\infty)}{1 - F_{ij}(\mathbf{0};\infty)}.
\end{equation}

In general, for any given profile of the initial numbers of cells, $\mathbf{N} = \{N_{ij}\}$, the extinction probability $P_E(t)$ at time $t$ is given by
\begin{equation}
P_E(t) = \prod_{i, j} F_{ij}(\mathbf{0}; t)^{N_{ij}}.
\end{equation}
The conditional probability $P_{noR}(t)$ is given by
\begin{equation}
P_{noR}(t) = \frac{ \prod_{i, j} F_{ij}(\mathbf{X}_{noR}; t)^{N_{ij}} -  \prod_{i, j} F_{ij}(\mathbf{0}; \infty)^{N_{ij}}}{1 - \prod_{i, j} F_{ij}(\mathbf{0}; \infty)^{N_{ij}}}.
\end{equation}

\subsection{Simple model} 
For the simple model with $2$ genotypes and $2$ compartments, we are able to derive closed-form approximations for special cases. The backward equations are as follows,
\begin{eqnarray}
\frac{\partial F_{00}}{\partial t} & = &  d_{00} + b_{00}(1-u) F_{00}^2 + b_{00} u F_{00} F_{10} + v F_{01} - (d_{00} + b_{00} + v) F_{00} \nonumber \\
\frac{\partial F_{01}}{\partial t} & = & d_{01} + b_{01}(1-u) F_{01}^2 + b_{01} u F_{01} F_{11} + v F_{00} - (d_{01} + b_{01} + v) F_{01} \nonumber \\
\frac{\partial F_{10}}{\partial t} & = & d_{10} + b_{10} F_{10}^2 + v F_{11} - (d_{10} + b_{10} + v) F_{10}  \nonumber \\
\frac{\partial F_{11}}{\partial t} & = & d_{11} + b_{11} F_{11}^2 +  v F_{10} - (d_{11} + b_{11} + v) F_{11}.
\end{eqnarray} 
These are coupled nonlinear Riccati equations, for which we do not have closed-form explicit solutions. 

For simplicity, let us further assume dichotomous drug environments: compartment $0$ is drug free, while compartment $1$ is distributed with a drug that inhibits cell growth. We have $d_{ij} = d_0$. In the drug-free compartment 0, both sensitive and resistant cells have supercritical replication potential in drug free compartment, but resistant cells have slightly lower replication rates than sensitive type due to the cost of the resistant mutation. We have $b_{00} > d_{00}$, $b_{10} > d_{10}$, and $b_{10} < b_{00}$. In the drug-containing compartment 1, sensitive cells have a subcritical replication potential while resistant cells still have a supercritical replication potential. We have $b_{01} < d_{01}$, $b_{11} > d_{11}$, and $b_{11} > b_{01}$.

For such a fitness landscape, the drug-free compartment provides far more favorable condition for breeding resistance than the drug-containing compartment. Therefore, the likely origin for treatment failure, these resistant cells that populate the drug-present compartment, should be attributed to the ``mutation-migration'' pathway than the ``migration-mutation'' pathway. In other words, it is less likely that resistance evolves \emph{in situ} in the drug-containing compartment than that resistance mutants are originated from the drug-free compartment and then migrate to the drug-containing compartment.  

\subsection{Drug-environment-dependent escape}

To verify these arguments above, let us first calculate and compare the likelihood of obtaining an escape mutation within one separate compartment (without any migration at all). For simplicity, the compartment is only seeded with a single sensitive cell. Denote by $b_0$ and $b_1$ the division rate of sensitive cells and resistant ones in that compartment. The death rate of the two types of cells is the same, $d_0 = d_1$. We have $b_0 > b_1>d_0$ for compartments with low concentrations, whereas $b_0 < d_0 < b_1$ for compartments with high concentrations. Let $f^0(x,y,t)$ denote the probability generating function for the lineages starting with a single sensitive cell, and let $f^1(x,y,t)$ denote the probability generating function for the lineages starting with a single resistant cell. We have the following differential equations:
\begin{eqnarray}
\frac{df^0}{dt} & = & d_0 + b_0(1-u) f^0 f^0 + b_0 u f^0 f^1 - (b_0 + d_0) f^0,\\
\frac{df^1}{dt} & = & d_1 + b_1 f^1 f^1  - (b_1 + d_1) f^1.
\end{eqnarray}
The initial condition is $f^0(x,y,0) = x$ and $f^1(x,y,0)=y$.

\subsubsection{$ b_0 > b_1>d_0$}

For sufficiently small mutation rates $u\ll 1$, we can derive a closed form approximation to the equations above. Let $r_i = b_i - d_i$ denote the net growth rate of type-$i$ cells, $i = 0, 1$. We have $r_0 > r_1 > 0$.
Neglecting the loss of sensitive cells due to rare mutations, we have 
\begin{equation}
\frac{df^0(x,t)}{dt} = d_0 + b_0 f^0 f^0 - (b_0 + d_0) f^0.
\end{equation}
This ODE above is explicitly soluble, and we obtain:
\begin{equation}
f^0(x,t) = \frac{b_0 x - d_0 - d_0(x-1)e^{r_0 t}}{b_0 x - d_0 - b_0(x-1)e^{r_0 t}}.
\end{equation}
Denote by $X_0(t)$ the number of sensitive cells at time $t$. Previous results show that $X_0(t)e^{-r_0 t}$ is a non-negative martingale and $X_0(t)e^{-r_0 t} \to W_0$ for $t \to \infty$. The Laplace transform of $W_0$ is given by
\begin{equation}
\mathbb{E} e^{-\theta W_0} = \mathbb{E} \left( e^{-\theta e^{-r_0 t}}\right)^{X_0} = f^0(e^{-\theta e^{-r_0 t}},t).
\end{equation}

For $t \to \infty$, using $e^{-\theta e^{-r_0 t}} - 1 \approx -\theta e^{-r_0 t}$, we obtain
\begin{equation}
\mathbb{E} e^{-\theta W_0}  \approx \frac{r_0 + d_0 \theta}{r_0 + b_0 \theta} = \frac{d_0}{b_0} + (1 - \frac{d_0}{b_0})\frac{1}{1 + \frac{b_0}{r_0} \theta}.
\end{equation}
Accordingly, we obtain the limiting distribution of $W_0$ for $t \to \infty$ as follows.
\begin{equation}
W_0 = \frac{d_0}{b_0} \delta_0  + (1 - \frac{d_0}{b_0}) \text{Exponential} (\frac{r_0}{b_0}),
\end{equation}
where $ \delta_0 $ is a point mass at zero (corresponding to extinction), and $\text{Exponential} (\frac{r_0}{b_0})$ is an exponential distribution with the exponent $\frac{r_0}{b_0}$. It is easy to see that, conditional on non-extinction, $X_0(t) e^{-r_0 t} \to V_0 = \text{Exponential} (\frac{r_0}{b_0})$. Thus the Laplace transform of $V_0$ is 
\begin{equation}
\mathbb{E} e^{-\theta V_0} = \frac{1}{1 + \frac{b_0}{r_0} \theta}.
\end{equation}

The rate at which resistant mutants arise is $u b_0 X_0 (t)$. Hence the conditional probability that the arrival time of the first mutant $\tau_1 > t$ is
\begin{eqnarray}
\mathbb{P}( \tau_1 > t | X_0(t) > 0) & = & \mathbb{E} e^{-\int_0^t u b_0 V_0 e^{r_0 s} ds} \nonumber \\
& = & \frac{1}{ 1 + \frac{u b_0^2}{r_0^2} (e^{r_0 t} - 1) } \approx  \frac{1}{ 1 + \frac{u b_0^2}{r_0^2} e^{r_0 t}  } .
\end{eqnarray}
The median arrival time 
\[
t_{\frac{1}{2}} = \frac{\log (r_0^2/u b_0^2)}{r_0}.
\]
We can see that $t_{\frac{1}{2}}$ is monotonically decreasing with increasing $b_0$ ($u \ll 1$). That is, the less harsh the compartment is ($b_0 \uparrow$), the sooner resistant mutants arise ($t_{\frac{1}{2}} \downarrow$). In this sense, compartments with lower drug concentrations are more likely the breeding ground for resistance, since the replication of sensitive cells is less affected than in compartments with higher drug concentrations. 

For small mutation rate $u\ll 1$, the escape probability $\mathbb{P}_1$ that a single sensitive cell establishes lineages without going extinction is approximately
\begin{equation}
\mathbb{P}_1 \approx \frac{r_0}{b_0} - u \frac{d_0 (b_0 d_1 - b_1d_0)}{b_0 b_1 r_0}.
\label{fenv}
\end{equation}

\subsubsection{$b_0 < d_0 < b_1$ }

In this unfavorable case, sensitive cells do not have a chance to survive but resistant cells do. The approximation method used above does not work here, since the branching process of sensitive cells is subcritical. To establish surviving lineages, sensitive cells must evolve resistant mutation right before quickly dying out. We cannot find a closed form for the arrival time of the first mutant for this case, but it is easy to see that if the mutant does arise (though with a lower chance), it takes much longer time due to the smaller birth rate than in benign environments. The escape probability of a single sensitive cell is approximately
\begin{equation}
\mathbb{P}_1 \approx - u \frac{b_0 r_1}{b_1 r_0},
\end{equation} 
which is much smaller than Eq.(\ref{fenv}). Moreover, we should note that the escape probability $\mathbb{P}_1$ is of the order $\sqrt{u}$ if sensitive cells are almost critical $|r_0| \to 0$:
\begin{equation}
\mathbb{P}_1 \approx \sqrt{u\frac{r_1}{b_1}}.
\end{equation} 

\subsection{Competing pathways to the emergence of resistance}

Returning to the simple model introduced above, we are now primarily concerned with which pathway leads faster to resistant cells that populate the drug-present compartment. To this end, we compare two distinct pathways. The migration-mutation pathway: sensitive cells first  
migrate from the drug-free compartment to the drug-present compartment, and then acquire resistance \emph{in situ}. The mutation-migration pathway: sensitive cells acquire resistant mutation in the drug-free compartment, followed by migration to the drug-present compartment. 

As shown above, we cannot obtain closed-from solutions for this four-type branching process.
To simplify our mathematical derivations, let us consider unidirectional emigration flow only from the drug-free compartment to the drug-present compartment. To compare which pathway is faster, let us further artificially separate these two pathways, making only one pathway at work at a time, instead of considering them simultaneously. In this way, each pathway can seen as a three-type branching process: 
\[
\text{type}\, 0 \xrightarrow{\mu_0} \text{type}\, 1 \xrightarrow{\mu_1} \text{type}\, 2.
\]
For the migration-mutation pathway, $\mu_0 = v$ and $\mu_1 = u b_{01}$; for the mutation-migration pathway, $\mu_0 = u b_{00}$ and $\mu_1 = v $. The difference of the fitness landscape between these two pathways lies in the intermediate type $1$, namely, the division rate $b_{01}$ of sensitive cells in the drug-present compartment versus the division rate $b_{10}$ of resistant cells in the drug-free compartment. It is therefore useful to first study a general three-type branching process and then obtain the results for each pathway by substituting the specific fitness landscape for each pathway. 

For rare mutations ($u\ll 1$) and low migration rates ($v \ll 1$), we are able to derive explicit closed-form approximations for the conditional probability of no type 2 cells by time $t$.
Let $b_i$ and $d_i$ be the birth and death rate of type $i$ cells, $i = 0, 1, 2$. Let $r_i = b_i - d_i$ be the net growth rate of type $i$ cells. We have $r_0 > 0$, $r_2 > 0$, and $r_1 < r_0$.
Conditional on non-extinction, the number of type 0 cells at time $t$ is given by 
\begin{equation}
X_0(t) = V_0 e^{r_0 t}
\end{equation}
Because type $1$ is less fit than type 0 ($r_1 < r_0$), we can approximate the number of type 1 cells at time $t$ as
\begin{equation}
X_1(t) = \mu_0 \int_0^t V_0 e^{r_0 s} e^{r_1(t-s)}ds =  \mu_0 V_0 \frac{e^{r_0 t}[1 - e^{(r_1 - r_0)t}]}{r_0 - r_1}
\sim \mu_0 V_0 \frac{e^{r_0 t}}{r_0 - r_1},
\end{equation}
where $x(t)\sim y(t)$ means $x(t)/y(t) \to 1$ for large $t$.

The conditional probability that there are no surviving type 2 cells by time $t$ is 
\begin{eqnarray} 
\mathbb{P} (X_2(t) = 0|X_0(t)>0) & = & \mathbb{E} \exp [-\mu_1 \int_0^t X_1(s) \frac{r_2}{b_2 - d_2 e^{-r_2(t - s)}} ds]\nonumber \\
& = & \mathbb{E} \exp [-\mu_0\mu_1 V_0 \int_0^t \frac{e^{r_0 s}}{r_0 - r_1} \frac{r_2}{b_2 - d_2 e^{-r_2(t - s)}} ds]\nonumber\\
&= & \frac{1}{1+ \theta_2(t) \frac{b_0}{r_0}},
\end{eqnarray}
where the integral $\theta_2(t) = \mu_0 \mu_1 \frac{r_2}{r_0 - r_1} \int_0^t \frac{e^{r_0s}}{b_2 - d_2 e^{-r_2 (t-s)}} ds$.

Substituting the fitness landscape of each pathway, we find a simple condition for the mutation-migration pathway faster than the migration-mutation pathway in leading to surviving resistant cells in the drug-containing compartment if the follow inequality holds:
\begin{equation}
\frac{b_{00}}{b_{00} - d_{00} - (b_{10} - d_{10})} > \frac{b_{01}}{b_{00} - d_{00} - (b_{01} - d_{01})}.
\label{mmc}
\end{equation}
Concerning cytostatic drugs that inhibits cell division, we have $d_{00} = d_{10} = d_{01} = d_{11} = d_0$, $b_{00} = b_0$, $b_{10} = (1 - s) b_0$ and $b_{01} = (1 - \delta) b_0$. $s$ denotes the fitness cost of resistant mutation in the absence of drugs, and $\delta$ denotes the fitness cost of sensitive cells in the presence of drugs. Then the above condition simplifies to 
\begin{equation}
\delta > \frac{s}{1 + s}.
\end{equation}
For cells in compartments with potent drugs and with sufficiently high drug concentrations, we have $\delta > s$. More generally, the emergence of resistance is more contingent on the mutation-migration pathway than the migration-mutation pathway, if the difference of drug concentration between two neighboring compartments leads to the landscape satisfying the inequality~(\ref{mmc}).

\section{Numerical results}
In what follows, we report the numerical results on the full model. To avoid boundary problems, we consider generic, symmetric spatial distributions of drug concentrations over compartments, according to a rescaled Normal distribution such that the degree of heterogeneity can be tuned by the variance of the distribution while keeping the total sum of concentrations constant. We compare the extinction probabilities and the time to resistance for sensitive cells initially located in each compartment, with varying migration rate and the spatial heterogeneity in drug concentrations.

The results are consistent with those reported in the main text (FigS.~\ref{lgm1} and \ref{lgm2}). Sanctuary sites are compartments with low drug concentration, which are responsible for producing the resistance that populates compartments with high drug concentration. Selection for resistance is more contingent on the mutation-migration pathway than the migration-mutation pathway, particularly in the presence of sanctuary sites. Moreover, only for small migration rates below a certain threshold does the spatial heterogeneity in drug concentrations help speed up the resistance evolution. However, excessively high migration rates actually slow down resistance emergence, as the role of compartment structure is diminished by frequent migration.

Furthermore, we observe interesting results when comparing local migration to global migration. Both the extinction probability and the time to resistance depend more sensitively on migration rate for global migration than for local migration (FigS.~\ref{lgm1}). This is because global migration favors the dissemination of resistance all over the place, whereas local migration constrains the migration within neighboring compartments with reduced drug difference such that accruing one point mutation is sufficient to offset the increasing drug concentration. As a consequence, resistance evolves much faster for local migration than global migration, particularly for sensitive cells located at the sanctuary sites with low drug concentrations. On the contrary, the escape probability of sensitive cells located in the compartments with high concentrations is hampered more by local migration than by global migration. In this case, local migration limits its dispersal range, while global migration increases the chance of moving to sanctuary sites thus leading to greater escape probabilities. More specifically, compared to global migration, local migration speeds up the evolution of resistance for sensitive cells initially located at sanctuary sites, while delaying the evolution of resistance for sensitive cells initially located at harsh compartments with high levels of drugs. Regarding the latter case, moreover, there exists an optimal intermediate global migration rate with which the time to resistance is minimal.

Sanctuary sites for sensitive cells result when varying from homogeneous drug distributions to increasingly heterogeneous drug concentrations across spatial compartments. It is obvious that compartments with low concentrations become the shelter for sensitive cells and thus provide benign condition for breeding resistance that can be selected for in compartments with excessively high concentrations (FigS.~\ref{lgm2}). Local migration takes less time to evolve resistance than global migration, if these cells are initially located at the sanctuaries. Opposite results are obtained if these cells are initially located at harsh compartments. Under local migration, there exists an optimal level of heterogeneity in drug concentrations that most delays the emergence of resistance. In contrast, increasing the spatial heterogeneity in drug concentrations helps prolong the time to resistance under global migration.
  
Altogether, these results confirm that our conclusions derived from the simple model in the main text are robust with respect to model extensions and varying model parameters. In addition, the scheme of migration plays an important role in the evolution of acquired resistance. However, future studies are needed in order to characterize, and eventually manipulate, the migration routes for metastatic cells. We think that constraining escape route of metastatic cells to sanctuary sites and limiting dissemination of evolved metastatic cells from sanctuary sites are vital to eliminate disseminated cancer.

% References for the SI
%\bibitem{Iwasa_JTB04}
%Iwasa Y, F Michor, MA Nowak (2004). Evolutionary dynamics of invasion and escape. J theor Biol 226 (2): 205-214.


\begin{thebibliography}{10}

\section*{REFERENCES}

\bibitem{Vogelstein_Science13}
Vogelstein B, et al. (2013) Cancer genome landscapes. Science 339(6127): 1546-1558.

\bibitem{Vogelstein_book}
Vogelstein B,  Kinzler KW (2002) The genetic basis of human cancer. Vol. 821. New York: McGraw-Hill.

\bibitem{Michor_NRC04}
Michor F, Iwasa Y, Nowak MA (2004) Dynamics of cancer progression. Nat Rev Cancer 4: 197-206. 

\bibitem{Nowak_PNAS04}
Nowak MA, Michor F, Komarova NL, Iwasa Y (2004) Dynamics of tumor suppressor gene inactivation. Proc Natl Acad Sci USA 101: 10635-10638. 

\bibitem{Bozic_PNAS10}
Bozic I, et al. (2010) Accumulation of driver and passenger mutations during tumor progression. Proc Natl Acad Sci USA 107: 18545-18550.

\bibitem{Sawyers_Nature04}
Sawyers C (2004) Targeted cancer therapy. Nature 432(7015): 294-297.

\bibitem{Druker_NEJM06}
Druker BJ, et al. (2006) Five-year follow-up of patients receiving imatinib for chronic myeloid leukemia. N Engl J Med 355: 2408-2417.

\bibitem{Sequist_JCO08}
Sequist LV, et al. (2008) First-line gefitinib in patients with advanced nonÐsmall-cell lung cancer harboring somatic EGFR mutations.  J Clin Oncol 26: 2442-2449.

\bibitem{Amado_JCO08}
Amado RG, et al. (2008) Wild-type KRAS is required for panitumumab efficacy in patients with metastatic colorectal cancer. J Clin Oncol 26: 1626-1634.

\bibitem{Chapman_NEJM11}
Chapman PB, et al. (2011) Improved survival with vemurafenib in melanoma with BRAF V600E mutation.  N Engl J Med 364: 2507-2516.

\bibitem{Bozic_Elife13}
Bozic I, et al. (2013) Evolutionary dynamics of cancer in response to targeted combination therapy. eLife 2: e00747.

\bibitem{Gottesman_ARM02}
Gottesman MM (2002) Mechanisms of cancer drug resistance. Annu Rev Med 53: 615-627.

\bibitem{Holohan_NRC13}
Holohan C, et al. (2013) Cancer drug resistance: an evolving paradigm. Nature Rev Cancer 13: 714-726.

\bibitem{Komarova_PNAS05}
Komarova NL, Wodarz D (2005) Drug resistance in cancer: principles of emergence and prevention. Proc Natl Acad Sci USA 102: 9714-9719.

\bibitem{Iwasa_Genetics06}
Iwasa Y, Nowak MA, Michor F (2006) Evolution of resistance during clonal expansion. Genetics 172: 2557-2566. 

\bibitem{Michor_CPD06}
Michor F, Nowak MA, Iwasa Y (2006) Evolution of resistance to cancer therapy. Curr Pharm Design 12: 261-271.

\bibitem{Durrett_TPB10}
Durrett R, Moseley S (2010) Evolution of resistance and progression to disease during clonal expansion of cancer. Theor Popul Biol 77: 42-48.

\bibitem{Danesh_JTB12}
Danesh K, Durrett R, Havrilesky LJ, Myers E (2012). A branching process model of ovarian cancer. J Theor Biol 314: 10-15.

\bibitem{Joyce_NRC08}
Joyce JA, Pollard JW (2009) Microenvironmental regulation of metastasis. Nature Rev Cancer 9: 239-252.

\bibitem{Sierra_DRU05}
Sierra A (2005) Metastases and their microenvironments: linking pathogenesis and therapy. Drug Resist Update 8: 247-257.

\bibitem{Quail_NatMed13}
Quail DF, Joyce JA (2013) Microenvironmental regulation of tumor progression and metastasis. Nat Med 19: 1423-1437.

\bibitem{Meeds_CCR08}
Meads MB, Hazlehurst LA, Dalton WS (2008) The bone marrow microenvironment as a tumor sanctuary and contributor to drug resistance. Clin Cancer Res 14: 2519-2526.

\bibitem{Tredan_NCI07}
Tr\'{e}dan O, et al. (2007) Drug resistance and the solid tumor microenvironment. J Natl Cancer Inst 99: 1441-1454.

\bibitem{Meads_NRC09}
Meads, MB, Gatenby RA, Dalton WS (2009) Environment-mediated drug resistance: a major contributor to minimal residual disease. Nat Rev Cancer 9: 665-674


\bibitem{Minchinton_NRC06}
Minchinton AI, Tannock IF (2006) Drug penetration in solid tumours. Nat Rev Cancer 6(8): 583-92


\bibitem{Zhang_Science11}
Zhang Q, et al. (2011) Acceleration of emergence of bacterial antibiotic resistance in connected microenvironments. Science 333: 1764-1767.

\bibitem{Greulich_PRL12}
Greulich P, Waclaw B, Allen RJ (2012) Mutational pathway determines whether drug gradients accelerate evolution of drug-resistant cells. Phys Rev Lett 109: 088101.


\bibitem{Hermsen_PNAS12}
Hermsen R, Deris JB, Hwa T (2012) On the rapidity of antibiotic resistance evolution facilitated by a concentration gradient.Proc Natl Acad Sci USA 109: 10775-10780.

\bibitem{Hallatschek_P12}
Hallatschek O (2012) Bacteria evolve to go against the grain. Physics 5: 93.

\bibitem{Pantel_NRCO09}
Pantel K,  Alix-Panabi\'{e}res C, Riethdorf S (2009) Cancer micrometastases. Nat Rev Clin Oncol 6: 339-351.


\bibitem{Chambers_NRC02}
Chambers AF, Groom AC, MacDonald IC (2002) Metastasis: dissemination and growth of cancer cells in metastatic sites. Nature Rev Cancer 2: 563-572.

\bibitem{Landen_JCO08}
Landen CN, Birrer MJ, Sood AK (2008) Early events in the pathogenesis of epithelial ovarian cancer. J Clin Oncol 26: 995-1005.

\bibitem{Klein_NRC09}
Klein CA (2009) Parallel progression of primary tumours and metastases. Nature Rev Cancer 9: 302-312.

\bibitem{Nguyen_NRC09}
Nguyen DX, Bos PD, Massagu\'{e} J (2009) Metastasis: from dissemination to organ-specific colonization. Nature Rev Cancer 9: 274-284.

\bibitem{Kang_CC13}
Kang YB, Pantel K (2013) Tumor Cell Dissemination: Emerging Biological Insights from Animal Models and Cancer Patients. Cancer Cell 23: 573-581.

\bibitem{Hanahan_Cell11}
Hanahan D, Weinberg RA (2011) Hallmarks of cancer: the next generation. Cell 144: 646-674.

\bibitem{Poste_Nature80}
Poste G, Fidler IJ (1980) The pathogenesis of cancer metastasis. Nature 283: 139-146.

\bibitem{Fidler_NRC03}
Fidler IJ (2003) The pathogenesis of cancer metastasis: the 'seed and soil' hypothesis revisited. Nature Rev Cancer 3: 453-458.

\bibitem{Pantel_JNCI99}
Pantel K,  Cote RJ, Fodstad {\O} (1999) Detection and clinical importance of micrometastatic disease. J Natl Cancer Inst  91: 1113-1124.

\bibitem{Pantel_NRC04}
Pantel K, Brakenhoff RH (2004) Dissecting the metastatic cascade. Nature Rev Cancer 4: 448-456.

\bibitem{Braun_NEJM05}
Braun S, et al.  (2005) A pooled analysis of bone marrow micrometastasis in breast cancer. N Engl J Med 353: 793-802.

\bibitem{Steeg_NM06}
Steeg PS (2006) Tumor metastasis: mechanistic insights and clinical challenges. Nat Med 12: 895-904.

\bibitem{Meng_CCR04}
Meng SD, et al. (2004) Circulating tumor cells in patients with breast cancer dormancy. Clin Cancer Res 10: 8152-8162.

\bibitem{Kerbel_JCB 94}
Kerbel RS, Kobayashi H, Graham CH (1994) Intrinsic or acquired drug resistance and metastasis: are they linked phenotypes? J Cell Biochem 56: 37-47.

\bibitem{Shtivelman_On97}
Shtivelman E (1997) A link between metastasis and resistance to apoptosis of variant small cell lung carcinoma. Oncogene 14: 2167-2173.

\bibitem{Glinsky_CL97}
Glinsky GV, et al. (1997) Apoptosis and metastasis: increased apoptosis resistance of metastatic cancer cells is associated with the profound deficiency of apoptosis execution mechanisms. Cancer Lett 115: 185-193.

\bibitem{Liang_CCDT02}
Liang Y, McDonnell S, Clynes M (2002) Examining the relationship between cancer invasion/metastasis and drug resistance. Curr Cancer Drug Targets 2: 257-277.

\bibitem{Smalley_MCT06}
Smalley KSM, et al. (2006) Multiple signaling pathways must be targeted to overcome drug resistance in cell lines derived from melanoma metastases. Mol Cancer Ther 5: 1136-1144.

\bibitem{Geyer_NEJM06}
Geyer CE, et al. (2006) Lapatinib plus capecitabine for HER2-positive advanced breast cancer. N Engl J Med 355: 2733-2743.

\bibitem{Sahai_NRC07}
Sahai E (2007) Illuminating the metastatic process. Nature Rev Cancer 7: 737-749.

\bibitem{Chaffer_Science11}
Chaffer CL, Weinberg RA (2011) A perspective on cancer cell metastasis. Science 331: 1559-1564.

\bibitem{Scott_RSI13}
Scott JG, et al. (2013) A mathematical model of tumour self-seeding reveals secondary metastatic deposits as drivers of primary tumour growth. J R Soc Interface 10: 20130011

\bibitem{Scott_NRC12}
Scott JG, Kuhn P,  Anderson ARA (2012) Unifying metastasisÑintegrating intravasation, circulation and end-organ colonization. Nature Rev Cancer 12: 445-446.

\bibitem{Wu_PNAS13}
Wu A, et al.  (2013) Cell motility and drug gradients in the emergence of resistance to chemotherapy. Proc Natl Acad Sci USA 110: 16103-16108.


\bibitem{Laird_BJC65}
Laird AK (1965) Dynamics of tumour growth: Comparison of growth rates and extrapolation of growth curve to one cell. Br J Cancer 19(2): 278.

\bibitem{Fournier}
Fournier DV, Weber E, Hoeffken W, Bauer M, Kubli F, Barth V (1980) Growth rate of 147 mammary carcinomas. Cancer 45(8): 2198-2207.

\bibitem{Friberg_JSO97}
Friberg S, Mattson S (1997) On the growth rates of human malignant tumors: Implications for medical decision-making. J Surg Oncol 65(4): 284-97.

\bibitem{Collins_TNM76}
Collins VP, Loeffler RK, Tivey H (1956) Observations on growth rates of human tumors. Am J Roentgenol Radium Ther Nucl Med 76(5): 988-1000.

\bibitem{Schwartz_Cancer61}
Schwartz M (1961) A biomathematical approach to clinical tumor growth. Cancer 14:1272-94.

\bibitem{Spratt_Cancer63}
Spratt JS, Spjut HJ, Roper CL (1963) The frequency distribution of the rates of growth and the estimated duration of primary pulmonary carcinomas. Cancer 16: 687-693.

\bibitem{Spratt_AS64}
Spratt Jr JS, Spratt TL (1964) Rates of growth of pulmonary metastases and host survival. Ann Surg 159(2): 161.

\bibitem{Steel_BJC66}
Steel GG, Lamerton LF (1966) The growth rate of human tumours. Br J Cancer 20(1): 74-86.

\bibitem{Komarova_PLOSONE09}
Komarova NL, Katouli AA, Wodarz D (2009) Combination of two but not three current targeted drugs can improve therapy of chronic myeloid leukemia. PLoS ONE 4(2): e4423.

\bibitem{Iwasa_PRSB03}
Iwasa Y, Michor F, Nowak MA (2003) Evolutionary dynamics of escape from biomedical intervention. Proc R Soc Lond Ser B Biol Sci 270: 2573-2578.

\bibitem{Foo_JTO12}
Foo J, Chmielecki J, Pao W, Michor F (2012) Effects of pharmacokinetic processes and varied dosing schedules on the dynamics of acquired resistance to erlotinib in EGFR-mutant lung cancer. J Thorac Oncol 7: 1583-1593.

\bibitem{Foo_JTB14}
Foo J, Michor F (2014) Evolution of acquired resistance to anti-cancer therapy. J Theor Biol 355: 10-20. 

\bibitem{Gatenby_CR09}
Gatenby RA, Silva AS, Gillies RJ, Frieden BR (2009) Adaptive therapy. Cancer Res 69: 4894-903. 


%Most targeted therapies have cytostatic effects
\bibitem{Leibel}
Hoppe R, Phillips TL, Roach III M (2010) Textbook of Radiation Oncology. United States: Saunders Elsevier (3rd Ed.)

\bibitem{Athreya_book}
Athreya KB, Ney PE (1972) Branching Processes. Berlin: Springer-Verlag.

\bibitem{Goldie_book}
Goldie JH, Coldman AJ (1998) Drug resistance in cancer: mechanisms and models. New York: Cambridge University Press.

\bibitem{Iwasa_JTB04}
Iwasa Y, Michor F, Nowak MA (2004) Evolutionary dynamics of invasion and escape.J Theor Biol 226(2): 205-214.

\bibitem{Haeno_Cell12}
Haeno H, et al. (2012). Computational modeling of pancreatic cancer reveals kinetics of metastasis suggesting optimum treatment strategies. Cell 148(1): 362-375.

\bibitem{Hermsen_PRL10}
Hermsen R, Hwa T (2010) Sources and sinks: a stochastic model of evolution in heterogeneous environments. Phys Rev Lett 105(24): 248104.




\bibitem{Baum_Lancet02}
Baum M, et al. (2002) Anastrozole alone or in combination with tamoxifen versus tamoxifen alone for adjuvant treatment of postmenopausal women with early breast cancer: first results of the ATAC randomised trial. Lancet 359: 2131-2139.

\bibitem{Feldmann_CR07}
Feldmann G, et al. (2007) Blockade of hedgehog signaling inhibits pancreatic cancer invasion and metastases: a new paradigm for combination therapy in solid cancers. Cancer Res 67: 2187-2196.



 
\bibitem{Bonhoeffer_Science04}
Bonhoeffer S, et al. (2004) Evidence for positive epistasis in HIV-1. Science 306: 1547-1550.


\bibitem{Woodhouse_Cancer97}
Woodhouse EC, Chuaqui RF, Liotta LA (1997) General mechanisms of metastasis. Cancer 80: 1529-1537.

\bibitem{Chambers_ACR00}
Chambers AF, et al. (2000) Clinical targets for anti-metastasis therapy. Adv Cancer Res 79: 91-121.

\bibitem{Wells_TPS13}
Wells A, et al. (2013) Targeting tumor cell motility as a strategy against invasion and metastasis. Trends Pharmacol Sci 34: 283Ð289

\bibitem{Michor_Nature05}
Michor F, et al. (2005) Dynamics of chronic myeloid leukaemia. Nature 435: 1267-1270.

\bibitem{Jordan_NEJM06}
Jordan CF,  Guzman ML, Noble M (2006) Cancer stem cells. N Engl J Med 355: 1253-1261.

\bibitem{Visvader_NRC08}
Visvader JE, Lindeman GJ (2008) Cancer stem cells in solid tumours: accumulating evidence and unresolved questions. Nature Rev Cancer 8: 755-768.

\bibitem{Zhou_JTB14}
Zhou D, Wang Y, Wu B (2014) A multi-phenotypic cancer model with cell plasticity. J Theor Biol 357: 35-45.

\bibitem{Komarova_PLOSONE07}
Komarova NL, Wodarz D (2007) Effect of cellular quiescence on the success of targeted CML therapy. PLoS ONE 2: e990.

\bibitem{Aguirre-Ghiso_NRC07}
Aguirre-Ghiso JA (2007) Models, mechanisms and clinical evidence for cancer dormancy. Nature Rev Cancer 7: 834-846.

\bibitem{Al-Hajj_PNAS03}
Al-Hajj M, et al. (2003) Prospective identification of tumorigenic breast cancer cells. Proc Natl Acad Sci USA 100: 3983-3988.

\bibitem{Wong_CR01}
Wong CW, et al. (2001) Apoptosis: an early event in metastatic inefficiency. Cancer Res 61: 333-338.

\bibitem{Luzzi_AJP98}
Luzzi KJ, et al. (1998) Multistep nature of metastatic inefficiency: dormancy of solitary cells after successful extravasation and limited survival of early micrometastases. Am J Pathol 153: 865-873.




%\bibitem{Delitala_JTB12}
%Delitala M, Lorenzi T (2012) A mathematical model for the dynamics of cancer hepatocytes under therapeutic actions. J Theor Biol 297: 88-102.

%\bibitem{Lorz_MMNA13}
%Lorz A, Lorenzi T, Hochberg ME, Clairambault J, Perthame B (2013) Populational adaptive evolution, chemotherapeutic resistance and multiple anti-cancer therapies. Mathematical Modelling and Numerical Analysis 47: 377-399.


% New refs...

%Tumor growth
%http://www.intechopen.com/books/advances-in-prostate-cancer/describing-prostate-cancer-dynamics-second-look-at-psa-doubling-time-and-psa-specific-growth-rate



% PDE models
\bibitem{Anderson_Cell06}
Anderson AR, Weaver AM, Cummings PT, Quaranta V (2006) Tumor morphology and phenotypic evolution driven by selective pressure from the microenvironment. Cell 127(5): 905-915.

\bibitem{Roose_SR07}
Roose T, Chapman SJ, Maini PK (2007). Mathematical models of avascular tumor growth. SIAM Review 49(2): 179-208.

\bibitem{Anderson_MMB05}
Anderson AR (2005) A hybrid mathematical model of solid tumour invasion: the importance of cell adhesion.  Math Med Biol 22(2): 163-186.

\bibitem{Delitala_JTB12}
Delitala M, Lorenzi T (2012) A mathematical model for the dynamics of cancer hepatocytes under therapeutic actions. J Theor Biol 297: 88-102.

\bibitem{Lorz_MMNA13}
Lorz A, Lorenzi T, Hochberg ME, Clairambault J, Perthame B (2013) Populational adaptive evolution, chemotherapeutic resistance and multiple anti-cancer therapies. Math Model Numer Anal 47: 377-399.

\bibitem{Thalhauser_BD10}
Thalhauser CJ, Lowengrub JS, Stupack D, Komarova NL (2010) Selection in spatial stochastic models of cancer: Migration as a key modulator of fitness. Biology Direct 5: 21.


%Drug concentration dependent killing
\bibitem{Regoes_AAC04}
Regoes RR, Wiuff C, Zappala RM, Garner KN, Baquero F, Levin BR (2004) Pharmacodynamic functions: a multiparameter approach to the design of antibiotic treatment regimens. Antimicrob Agents Chemother 48(10): 3670-3676.

\bibitem{Hill_IND94}
Hill BT, Whelan RD, Shellard SA, McClean S, Hosking LK (1994) Differential cytotoxic effects of docetaxel in a range of mammalian tumor cell lines and certain drug resistant sublinesin vitro. Invest New Drugs 12(3): 169-182.

\bibitem{Goutelle_FCP08}
Goutelle S, Maurin M, Rougier F, Barbaut X, Bourguignon L, Ducher M, Maire P (2008) The Hill equation: a review of its capabilities in pharmacological modelling. Fundam Clin Pharmacol 22(6): 633-648.

\bibitem{Hill_IJC87}
Hill BT, Whelan RD, Gibby EM, Hosking LK, Thomas Rupniak H, Sheer D, Shellard SA (1987) Establishment and characterisation of three new human ovarian carcinoma cell lines and initial evaluation of their potential in experimental chemotherapy studies. Int J Cancer 39(2): 219-225.

%Fletcher JI, Haber M, Henderson MJ, Norris MD (2010) ABC transporters in cancer: more than just drug efflux pumps. Nat Rev Cancer 10(2):147-156
%Gottesman MM, Fojo T, Bates SE (2002) Multidrug resistance in cancer: role of ATP-dependent transporters. Nat Rev Cancer 2(1):48-58
%Fodale V, Pierobon M, Liotta L, Petricoin E (2011) Mechanism of cell adaptation: when and how do cancer cells develop chemoresistance? Cancer J. 17(2):89-95
%Kavallaris M (2010) Microtubules and resistance to tubulin-binding agents. Nat Rev Cancer 10(3):194-204
%Dean, Michael, Tito Fojo, and Susan Bates. "Tumour stem cells and drug resistance." Nature Reviews Cancer 5.4 (2005): 275-284.

\bibitem{Rosenbloom_NM12}
Rosenbloom DIS, Hill AL, Rabi SA, Siliciano RF, Nowak MA (2012) Antiretroviral dynamics determines HIV evolution and predicts therapy outcome. Nat Med 18: 1378-1385.

\bibitem{Antal_JSM10}
Antal T, Krapivsky PL (2010) Exact solution of a two-type branching process: Clone size distribution in cell division kinetics. J Stat Mech 2010: P07028. 

\bibitem{Conway_PLoSCB11}
Conway JM, Coombs D (2011) A Stochastic model of latently infected cell reactivation and viral blip generation in treated HIV patients. PLoS Comput Biol 7: e1002033.

\end{thebibliography}
\end{document}